\documentclass[preprint,12pt,authoryear]{elsarticle}

\usepackage{amssymb}
\usepackage{enumitem}
\usepackage{booktabs} 
\usepackage{threeparttable}
\usepackage{multirow}
\usepackage{algorithm}
\usepackage{algpseudocode}
\usepackage{subcaption}
\usepackage{comment}
\usepackage{units}
\usepackage{amsmath}
\usepackage{soul}
\usepackage[dvipsnames]{xcolor}

\usepackage{xcolor}

\journal{arXiv.org}

\begin{document}
\begin{frontmatter}



\title{Leader-Follower Identification with Vehicle-Following Calibration for Non-Lane-Based Traffic}


\author[inst1]{Mihir Mandar Kulkarni\fnref{eqt}}

\affiliation[inst1]{organization={Zachry Department of Civil \& Environmental Engineering, Texas A\&M University},
            postcode={3136}, 
            addressline={TAMU}, 
            city={College Station},
            state={TX},
            country={USA}}

\author[inst2]{Ankit Anil Chaudhari\corref{cor1}\fnref{eqt}}
\affiliation[inst2]{organization={Chair of Econometrics and Statistics esp. in the Transport Sector, Institute of Transport and Economics},
            addressline={Faculty of Transport and Traffic Sciences, Technische Universität Dresden}, 
            postcode={01187},
            city={Dresden},
            country={Germany}}
\ead{ankit_anil.chaudhari@tu-dresden.de}
\cortext[cor1]{Corresponding Author}
\fntext[eqt]{These authors have contributed equally to this work}
\author[inst3]{Karthik K. Srinivasan}
\author[inst3]{Bhargava Rama Chilukuri}
\author[inst2]{Martin Treiber}            
\author[inst2,inst4]{Ostap Okhrin}

\affiliation[inst3]{organization={Department of Civil Engineering},
            addressline={Indian Institute of Technology Madras}, 
            city={Chennai},
            postcode={600036}, 
            country={India}}
\affiliation[inst4]{organization={Center for Scalable Data Analytics and Artificial Intelligence (ScaDS.AI)},
            city={Dresden/Leipzig},
            country={Germany}}
\begin{abstract}
Most car-following models were originally developed for lane-based traffic. Over the past two decades, efforts have been made to calibrate car-following models for non-lane-based traffic. However, traffic conditions with varying vehicle dimensions, intermittent following, and multiple leaders often occur and make subjective Leader-Follower (LF) pair identification challenging. In this study, we analyze Vehicle Following (VF) behavior in traffic with a lack of lane discipline using high-resolution microscopic trajectory data collected in Chennai, India. The paper’s main contributions are threefold. Firstly, three criteria are used to identify LF pairs {from the driver's perspective}, taking into account the intermittent following, lack of lane discipline due to consideration of lateral separation, and the presence of in-between vehicles. Second, the psycho-physical concept of the regime in the Wiedemann-99 model is leveraged to determine the traffic-dependent "influence zone" for LF identification. Third, a joint and consistent framework is proposed for identifying LF pairs and estimating VF parameters. The proposed methodology outperforms other heuristic-based LF identification methods from the literature in terms of quantitative and qualitative performance measures. The proposed approach can enable robust and more realistic LF identification and VF parameter calibration with practical applications such as LOS analysis, capacity, and travel time estimation. 
\end{abstract}




\end{frontmatter}


\section{Introduction}
\label{sec:Intro}
Microscopic traffic simulation models are widely used to predict traffic performance measures like delays \citep{tian2002variations}, level of service \citep{geistefeldt2014assessment}, or vehicular emissions \citep{abou2013using}. Traffic simulations are based on several models, including car-following and lane-changing models, in order to represent the lateral and longitudinal movement and placement of vehicles in the traffic stream. Vehicle following behavior refers to how a subject vehicle reacts to the motion of its leader (the vehicle immediately ahead) when they are sufficiently close to each other such that the leader has a considerable influence on the follower.

Traffic conditions can be classified based on the presence or absence of lane discipline into the following two categories: Lane-Based (LB) flow and Non-Lane-Based (NLB) flow (both categories may include heterogeneous traffic, i.e., several types of vehicles). Compared to LB traffic, which has received the bulk of the attention in existing literature \citep{gazis1961nonlinear,wiedemann1974simulation,gipps1981behavioural, bando1998analysis, Treiber2000}, analyzing car-following behavior and identification of LF pairs is more challenging in NLB traffic flow for a number of reasons. The interaction among different vehicle types is complex, and the influence of surrounding vehicles beyond the assumed leader and follower pair must be considered. Moreover, the movements of smaller vehicles like two-wheelers and their partial presence in the influence region between the leader and follower pair can make the car-following process intermittent rather than a continuous one. Furthermore, the lack of lane discipline and varying vehicle sizes necessitates considering the role of lateral overlap in car-following behavior. For these reasons, methods for identifying LF pairs based on lane-based traffic are unlikely to be effective for NLB streams. To address these challenges, this paper develops a joint and consistent methodology for LF identification and car-following model parameter calibration for such NLB traffic flows.

The methodology in the NLB context is illustrated in this paper using trajectory data from a divided arterial in Chennai City, India. Trajectory data offers several advantages for the calibration of car-following models. 

First, it can provide information about the complete set of driving regimes, enabling a thorough understanding of driving behavior \citep{sharma2019more}. As a result, it is shown that trajectory data can be used for more accurate and unbiased calibration \citep{anil2022calibrating}. Further, it provides richer and more extensive information than cross-sectional and probe-based data \citep{sekula2018estimating}. {Moreover, it provides complete information on all trajectories in a certain spatiotemporal region, which is a prerequisite for determining the relevant vehicles in the local neighborhood.} Finally, the variety and quantity of such data enable the development of data-driven and machine learning-based models \citep{papathanasopoulou2015towards}, which may be more accurate for forecasting purposes. For these reasons, there is a growing use of such data in developing and calibrating \citep{kesting2008calibrating} different car-following models such as the Wiedemann-99 model \citep{wiedemann1974simulation}, the IDM \citep{Treiber2000}, the OVM \citep{bando1998analysis} and the Gipps car-following model \citep{gipps1981behavioural}. As this study is focused on different types of leaders and followers, we are using the {more general term \textit{vehicle following (VF)} instead of car following}. An important step in VF model parameter estimation in many models is the identification of leader-follower pairs. For this purpose, different studies \citep{zhu2018modeling,chong2013rule,leblanc2013longitudinal,higgs2013two,durrani2016calibrating} use a variety of heuristics in the literature which are based on some strong assumptions and restrictions discussed in the next section.

This research is motivated by the following considerations. First, the methods for LF identification vary widely across the literature and are based on heuristics involving trial and error, subjectivity, or the use of arbitrary thresholds without calibration, e.g., for the longitudinal clearance or the headway. As a result, it is difficult to transfer and compare parameters across different studies. For instance, \citet{brockfeld2004calibration} used all vehicle pairs in calibrating model parameters, which can lead to an unrepresentative sample of the following behavior. Recognizing this limitation, \cite{anand2019calibration} use simple heuristics for LF identification but are based solely on spatial or temporal proximity (e.g., distance or headway thresholds). Two main limitations of the heuristic approaches for LF identification include: 1. choice of thresholds is arbitrary, and 2. trial and error procedures may yield erroneous LF pairs, affecting and potentially invalidating the subsequent estimation of the VF parameters. 

The second main motivation is that there are several reasons that mere physical proximity (or its absence) may not always be an adequate indicator of the influence of the leader’s motion on a subject vehicle’s behavior. Due to differences in the vehicles (age,  dimensions, etc.), the relative speed between leader and follower, or driver characteristics (aggressiveness, skill, etc.), the response to the same physical or temporal gap may be different and thus inadequate to demarcate following vs. non-following regimes. Furthermore, proximity-based measures do not recognize the possibility of different driving regimes, including some non-following episodes within the same gap or time headway. Besides, the following process may be discontinuous due to psychophysical constraints such as visual angle or its rate of change. 

 Third, existing methods for LF identification do not adequately account for non-lane-based traffic features. For instance, the criteria of being in the same lane is not properly defined because of staggered following or simply because there are no lanes. Further, it is not easy to identify whether intervening vehicles are present between a potential LF pair due to these varying spatial configurations and their partial presence in the influence area, as illustrated in Section 3. The presence and varying influence of other surrounding vehicles between the subject vehicle and the staggered leader can also lead to intermittent following in the NLB traffic context. 
 
To address these gaps and considerations, the following objectives are pursued in this paper:
\begin{enumerate}[label=(\alph*)]

    \item Demonstrate the limitations of existing heuristics for LF identification in mixed traffic.
    \item Develop a new methodology for LF identification in mixed traffic that incorporates

    \begin{enumerate}[label=\roman*.]
        \item joint and consistent procedure for LF identification and vehicle-following parameters estimation,
        \item information about psychophysical driving regimes,
        \item and more accurate representation of heterogeneous traffic such as the presence of potential surrounding intervening vehicles and lateral separation between the leader and follower.
    \end{enumerate}
    \item Evaluate the proposed approach in comparison with the existing LF identification methods.
\end{enumerate}
{Using our new methodology, we show that it outperforms existing heuristics in terms of the accuracy of predicted trajectories.}

The structure of the paper is as follows. Section \ref{sec: Lit Rev} summarizes various criteria used in the literature to identify LF pairs in homogeneous and heterogeneous traffic and their associated advantages and disadvantages. Section \ref{sec: data} outlines the data utilized in the study and preliminary analysis to illustrate the limitations of existing LF identification methods. Section 4 provides the rationale and an elaborate explanation of the proposed methodology. In Section 5, the salient findings and results from this study are presented, including an evaluation of the proposed method in relation to other LF identification methods. Section 6 presents some concluding remarks along with directions for future work.

\section{Literature Review}
\label{sec: Lit Rev}
A comprehensive review of car-following literature is provided in \cite{olstam2004comparison}. 
To analyze vehicle-following behavior using trajectory data, it is crucial to identify 
the points of the \emph{influence zone} for the assumed leader-follower pair where the follower’s motion is influenced by the leader. Various methods have been used to identify influence {zones}, ranging from observations of traffic videos to measuring lagged acceleration correlations between an assumed leader and an assumed follower. Heuristic thresholds on longitudinal clearance, relative speed, lateral overlap, and time headways have been widely applied to identify {the influence zone}. These thresholds depend on the type of traffic and method of data collection. For example, verifying that the assumed leader and follower are traveling in the same lane may be sufficient for LB traffic but not for NLB traffic because of the lack of lane discipline. If the data is collected by a {range} sensor on the following vehicle {detecting only vehicles straight ahead}, it may be sufficient to ensure that the leader remains in the same lane \citep{higgs2013two}. If the data, however, contains all trajectories (e.g., the NGSIM \citep{us2008ngsim} or Chennai data \citep{kanagaraj2015trajectory}), then it becomes crucial to identify periods {where a given leader-follower pair is not interrupted by other surrounding vehicles.}

There is a growing interest in developing vehicle-following models for NLB traffic in the last decade \citep{mathew2011car,kanagaraj2013evaluation,das2019multivariate,adavikottu2023modelling,raveendran2024modeling}. Some studies have tried to include lateral gaps and movement in the analysis \citep{gaddam2020two,kanagaraj2018self,sharath2020enhanced,nirmale2024two}. A variety of VF models have been used, but only a few have included psycho-physical driving regime information \citep{raju2021modeling,anil2022calibrating} in the model.

Many of these studies have not explicitly dealt with LF pair identification for various reasons.  Some studies have used synthetic data \citep{ravishankar2011vehicle,das2019multivariate} or simulated data \citep{mathew2010calibration} instead of naturalistic data. Other studies on NLB have used aggregated macroscopic measures to calibrate the vehicle-following parameters \citep{mathew2010calibration,asaithambi2018study}, and hence, LF identification was not examined sufficiently. A few researchers have used heuristics such as aggregate hysteresis for LF identification \citep{raju2021modeling, das2019multivariate}. However, the comprehensive treatment of LF and VF simultaneously is underexplored.

\begin{table}[!ht]
    \centering
    \caption{{Criteria and thresholds for the influence zone from existing literature. If several criteria from the same source are listed, all need to be true for a valid influence zone.} 
    }
    \resizebox{\textwidth}{!}
    {
    \begin{tabular}{p{4.5cm}p{6cm}p{2cm}p{2.5cm}}
    \hline
        \textbf{Literature} & \textbf{Criteria} & \textbf{Threshold value} &\textbf{Remarks} \\ \hline
        \cite{zhu2018modeling} & Longitudinal clearance & $<\unit[120]{m}$ & \multirow{2}{3cm}{Expressway in China}  \\ 
        & Lateral distance between center of follower and radar target & $<\unit[2.5]{m}$ & ~\\ 
        \cite{sun2021modeling} & Longitudinal clearance & $>\unit[7]{m}$ & NGSIM data \\ 
         & Longitudinal clearance & $<\unit[120]{m}$ & ~\\
         & Relative speed & $<\unit[2.5]{m/s}$ & ~\\
         & Speed of follower & $>\unit[5]{m/s}$ & ~\\
         & Correlation between acceleration of follower and speed difference & $>0.6$& ~\\ 
        \cite{anand2019calibration} & Longitudinal clearance & $<\unit[30]{m}$ & \multirow{4}{1em}{NLB Traffic} \\ 
        &Lateral displacement with leader& $<\unit[3]{m}$ \\
        & Headway & $<\unit[2]{s}$ & ~\\ 
         & Continuous following & $>\unit[5]{s}$ & ~\\ 
        \cite{fernandez2011secondary} & Longitudinal clearance & $<\unit[100]{m}$ & \multirow{2}{3cm}{Urban arterial roads}\\ 
        & Length of following section & $>\unit[100]{m}$&~ \\ 
        \cite{chong2013rule} & Lateral clearance & $<\unit[1.9]{m}$ &~\\ 
        \cite{raju2021modeling} & Lateral clearance & $<\unit[1.5]{m}$ & NLB traffic \\ 
        \cite{papathanasopoulou2018flexible} & Lateral safety distance of both sides & $<\unit[0.2]{m}$\\ 
        \cite{nagahama2021detection} & Edge to edge lateral clearance & $<\unit[0.3]{m}$ & \\ 
        \cite{leblanc2013longitudinal}& Relative speed & $<\pm\unit[2]{m/s}$&~\\ 
         & Follower's Speed & $>\unit[11.2]{m/s}$&~\\ 
         & Continuous following & $>\unit[15]{s}$&~ \\ 
        \cite{higgs2013two} & Follower's Speed & $>\unit[5.55]{m/s}$ & To minimise effect of traffic jam\\ 
        &  Length of vehicle following period & $>\unit[30]{s}$ & \multirow{3}{3cm}{For trucks}\\ 
         & Longitudinal clearance & $>\unit[61]{m}$&~ \\ 
         & Longitudinal clearance & $<\unit[120]{m}$&~\\ 
        \hline
    
     \end{tabular}
     }
    \label{tab:Literature Review}
\end{table}
Table \ref{tab:Literature Review} gives an overview of different LF identification criteria used in literature, which  are explained in detail as follows. 
Longitudinal clearance between the assumed leader and follower has been widely used to determine the influence zone by \citep{zhu2018modeling,sun2021modeling,anand2019calibration,higgs2013two,fernandez2011secondary}. Lateral clearance or lateral overlapping width between leader and follower is another criterion to identify the influence zone \citep{zhu2018modeling,chong2013rule,raju2021modeling,papathanasopoulou2018flexible,nagahama2021detection}. In lane-based traffic, it could be sufficient to ensure that the assumed leader and follower are in the same lane \citep{leblanc2013longitudinal, sun2021modeling}. Variables like relative speed (follower speed minus leader speed, i.e., approaching rate) \citep{leblanc2013longitudinal,sun2021modeling}, speed of the follower \citep{leblanc2013longitudinal,higgs2013two,sun2021modeling}, and time headway \citep{anand2019calibration} have also been used to determine the influence zone.\

After classifying individual trajectory points {as falling within or outside the influence zone}, the pair should be classified as an LF or non-LF pair. The following duration  \citep{anand2019calibration,leblanc2013longitudinal,higgs2013two} and length of the following segment \citep{fernandez2011secondary} are used to identify whether the pair can be recognized as LF or non-LF. \cite{sun2021modeling} used an additional condition based on the Pearson correlation. \cite{raju2021modeling,das2019multivariate} used hysteresis plots to determine leader-follower pairs.
To identify LF pairs in non-lane-based traffic, it is crucial to consider the effect of other surrounding vehicles between the assumed leader and follower. This situation may not arise in lane-based traffic but can frequently happen in non-lane-based traffic due to the varying dimensions of vehicles. \cite{raju2021modeling} used space-time plots to identify other surrounding vehicles that could influence the interaction between the assumed leader-follower pair. Unfortunately, visual observation of space-time plots is not scalable if the data is large.
Moreover, if an intermediate vehicle is intermittently present for more time between potential LF pair, it may not be practical to consider the whole potential pair for calibration. \cite{nagahama2021detection} used Voronoi-triangulation-based logic to detect the influence of surrounding vehicles. According to them, if an edge of a triangle exists between the leader and follower, it serves as one of the influencing conditions. \cite{sharma2019modelling} generated different trajectory groups and studied the impact of trajectory .\cite{chen2020investigating} incorporate human factors in driving modeling by identifying long-term and short-term driving characteristics and calibrating the vehicle following models for those characteristics; these driving characteristics identification is similar to influence points identification as proposed in Section 4.1.

The above literature on the identification of LF pairs reveals several gaps:
1) Using a fixed threshold for longitudinal clearance may result in inaccurate identification of the influence zone.
2) Lateral clear gap or overlap thresholds may not be the same for all vehicle types. For instance, a \unit[0.2]{m} overlap might influence a two-wheeler following a vehicle but not necessarily a truck.
3) Existing literature relies on heuristic-based thresholds for the duration of the continuous following, with limited consideration for variations in traffic conditions; long following durations can often be observed in LB traffic because of the lane discipline, but for NLB traffic, intermittent following is frequent, \citep{anil2022calibrating}, also as shown in Section 3.2.
4) Hysteresis plots are not easily scalable, and they are subjective.
5) Previous works do not comprehensively address the identification of other potentially influencing vehicles at every point in the trajectory.

The next section explains the data used for the study and some preliminary analysis of the existing LF identification methods.

\section{Data Description and Preliminary Analysis}
\label{sec: data}
\subsection{Data Description}
\begin{figure}[hbt!]
  \centering
  \includegraphics[width=1\textwidth]{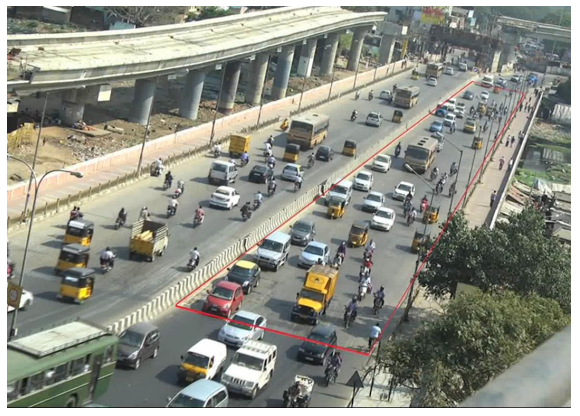}
  \caption{Study Section in Chennai, India \citep{kanagaraj2015trajectory}}\label{Fig:Studysection}
\end{figure}

This study {is based on video data with two frames per second} collected from a six-lane divided urban arterial in Chennai, India, as detailed in \cite{kanagaraj2015trajectory}. The selected study section, depicted in Figure \ref{Fig:Studysection}, is situated on a bridge, ensuring a uniform and consistent roadway segment for analysis. In the vicinity, there are no immediate intersections, bus stops, parked vehicles, or any other factors that could potentially influence {the drivers' behavior}. Additionally, there is no interaction between vehicle traffic and pedestrians. The width of the section under study {is \unit[11.2]{m} and its length \unit[245]{m}}. The video data was recorded from 10:00 AM to 3:30 PM. The coordinates, dimensions, and class of all vehicles in the video sequences during 30 min between 2:45 PM and 3:15 PM were obtained using a trajectory extractor. The data consists of 3016 vehicles with six different vehicle classes and 130,137 data points. {The vehicle population} consist of 1703 (57$\%$) motorized two-wheelers, 802 (27\%) cars, 367 (12\%) three-wheelers, 95 (3\%) buses, 40 (1\%) light trucks, and 9 (0.29\%) heavy trucks. 
The study area has three nominal lanes, which are largely ignored. 

From the trajectory data, we selected 2078 potential LF pairs using the influence area method proposed by \cite{anand2019calibration} (cf. Table~\ref{tab:Literature Review}).
These pairs consist of 1099 motorized two-wheelers, 623 cars, 249 three-wheelers, 59 buses, 42 light trucks, and 6 heavy trucks as follower vehicles. This work focuses on the 623 potential LF pairs with a car following.
For each pair, we extracted the longest continuous duration where the longitudinal clearance between the assumed leader and the assumed follower remained positive. These extracted parts of trajectories are used for further analysis to ensure that the potential vehicle-following periods are not discontinuous.

\subsection{{Intervening surrounding vehicles}}

\begin{figure}[hbt!]
  \centering
  \includegraphics[width=1\textwidth]{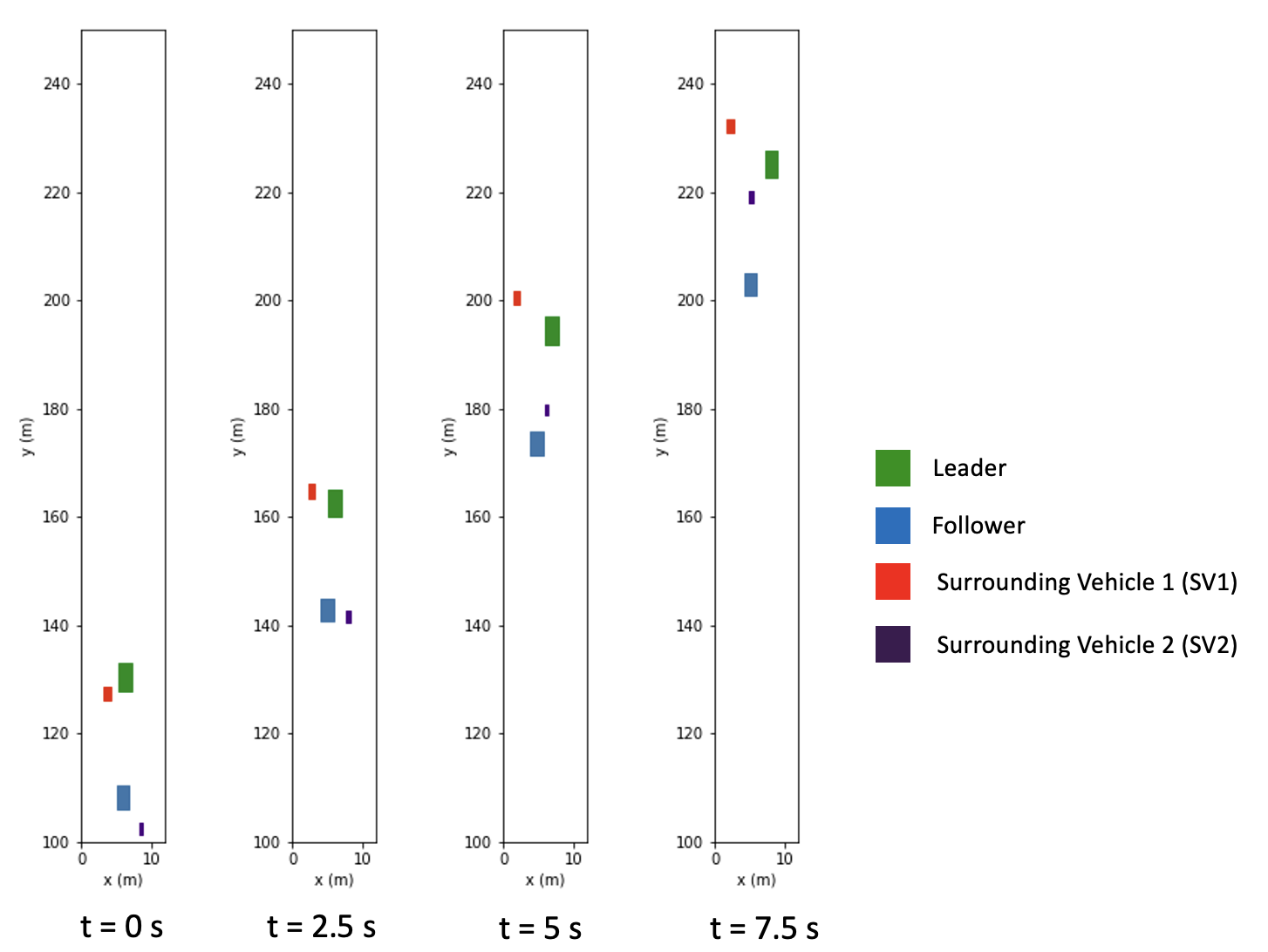}
  \caption{Trajectory of a potential LF pair with surrounding vehicles and intermittent following.}\label{Fig:Trajectorires_potential_5}
\end{figure}

Figure \ref{Fig:Trajectorires_potential_5} shows an example trajectory of a potential LF pair in the data. At $t$ = \unit[0]{s}, the red surrounding vehicle (SV1) is present between the potential leader (Green) and follower (Blue), but it may not be influencing the LF interaction. The SV1 moves away, and at $t$ = \unit[2.5]{s}, there is no surrounding vehicle between the leader and follower. But in the next few seconds, another vehicle (SV2) cuts in. From $t$ = \unit[5]{s} to $t$ = \unit[7.5]{s}, we can say that the SV2 (purple) influences the leader-follower interaction. Intervening surrounding vehicles are often observed in the NLB traffic. However, no methods in the literature look at the potential intervening vehicles at each point of the trajectory. 

One of the goals of this work is to evaluate existing LF identification methods. Limitations of the existing methods are pointed out in the following section.

\subsection{Limitation of existing LF identification methods}
Existing LF identification methods are evaluated based on the procedure of pair identification and the performance of identified LF pairs for calibration of the Wiedemann-99 model. Existing methods are as follows:

\begin{enumerate}[label=(\alph*)]
    \item 	LF identification based on heuristic thresholds of longitudinal clearance and time headway \citep{fernandez2011secondary,zhu2018modeling,sun2021modeling,anand2019calibration}
    \item LF identification based on hysteresis plots \citep{raju2021modeling}
    
\end{enumerate}

\subsubsection{Limitations on heuristic thresholds based methods}

\begin{figure}[!htpb]
 \begin{subfigure}{0.9\textwidth}
  \centering
  \includegraphics[width=1\textwidth]{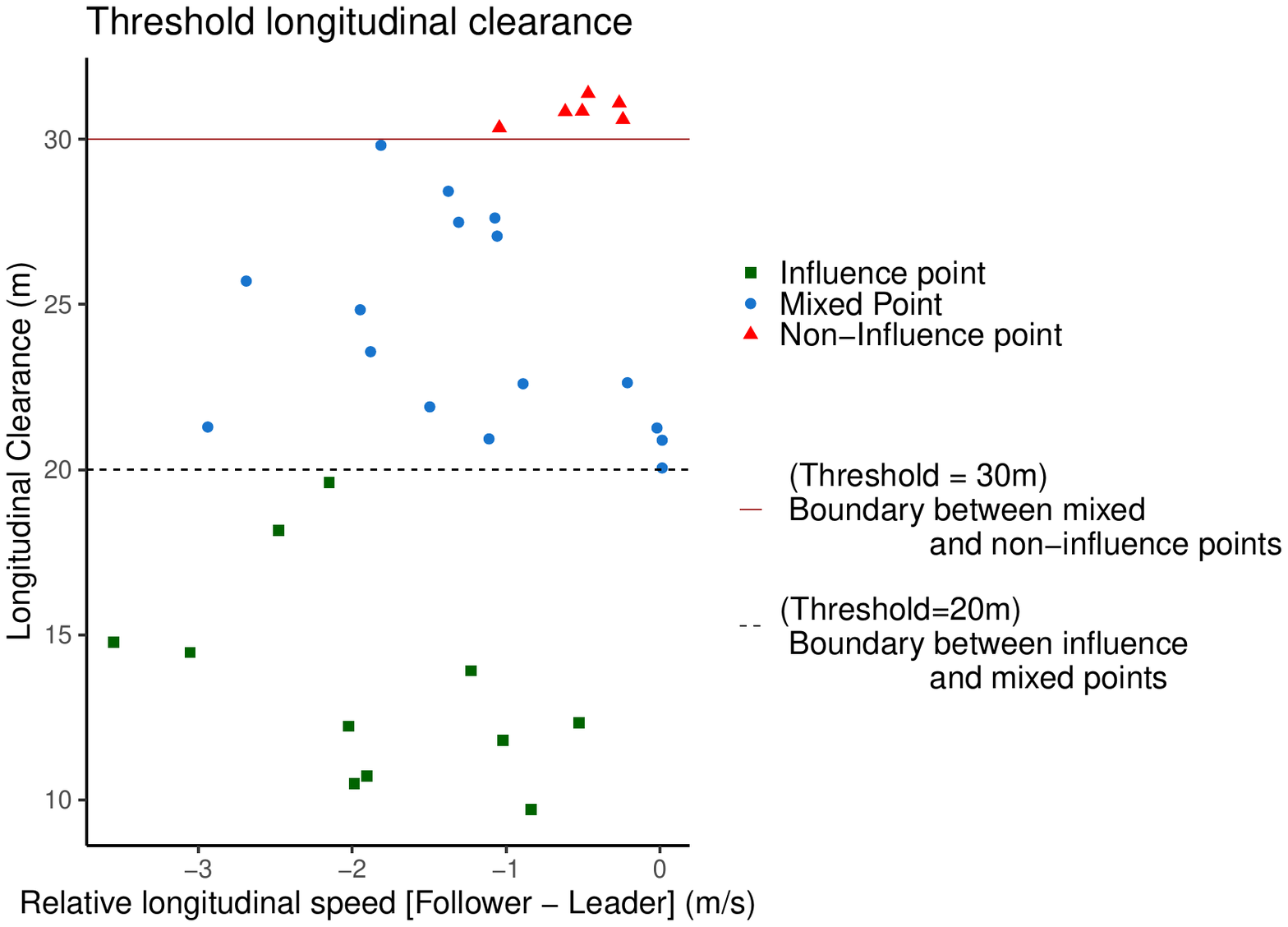} 
  \caption{}
  \end{subfigure}
  \begin{subfigure}{0.9\textwidth}
  \includegraphics[width=1\textwidth]{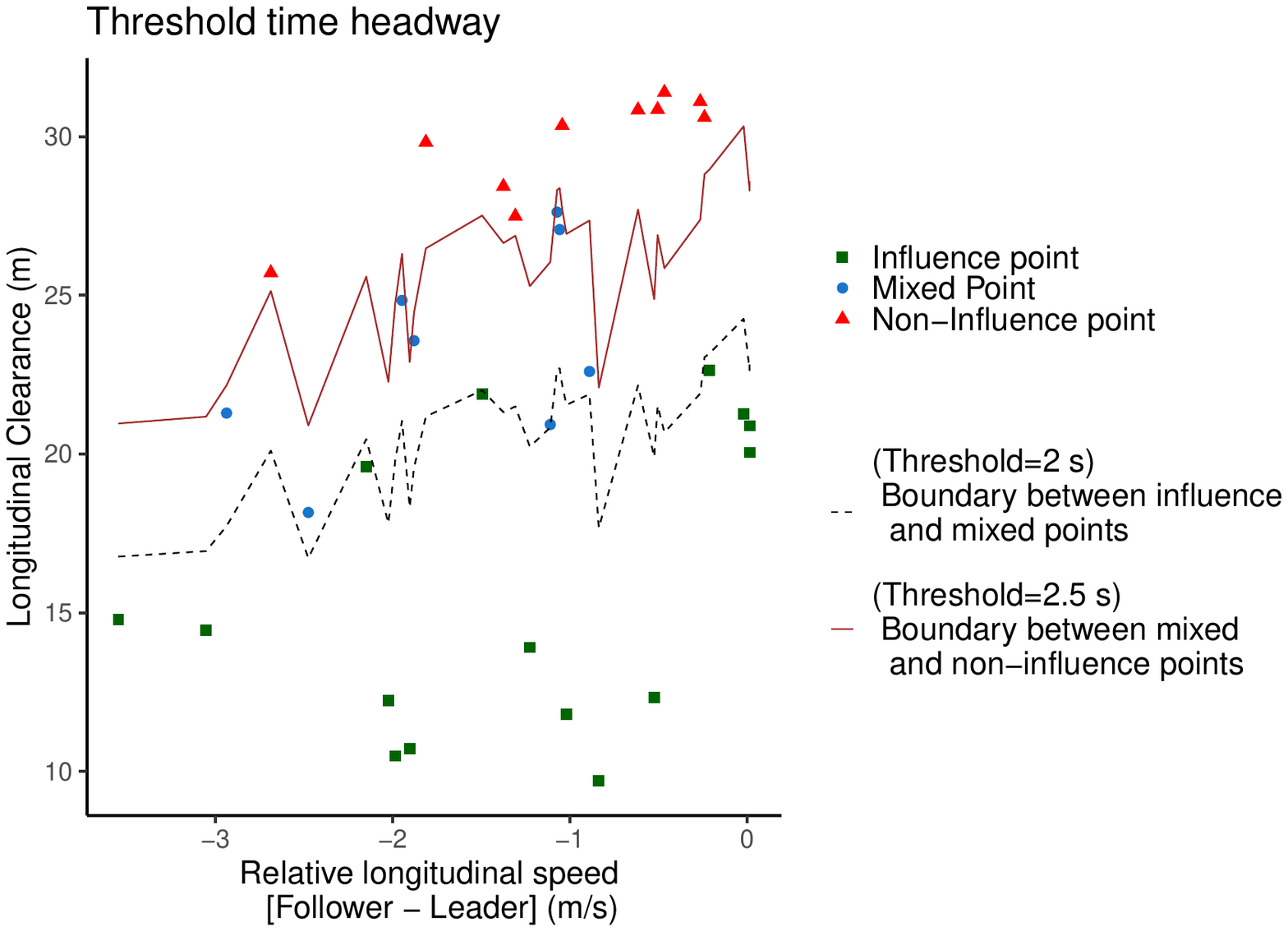} 
   \caption{}

  \end{subfigure}
  \caption{Identification of {the influence zone} for one LF pair using a) Longitudinal clearance threshold b) Headway threshold.}\label{Fig:InflueNonInfleTheshold_4}
\end{figure}

It is crucial to critically evaluate existing LF identification methods based on physical thresholds, as changes in thresholds can have significant implications for influence point determination. Figure \ref{Fig:InflueNonInfleTheshold_4} shows the effect of changing thresholds for two of the heuristic methods. In the first method, longitudinal clearance between leader and follower is used to classify points into influencing or non-influencing.
However, when the longitudinal clearance threshold is changed from \unit[20]{s} to \unit[30]{m}, a considerable number of points in blue got reclassified as influence points (Fig. \ref{Fig:InflueNonInfleTheshold_4}(a)). We call the reclassified points a mixed points. The same ambiguity can be observed in the second method, where a time gap threshold is used to classify points into influencing or non-influencing (Fig.\ref{Fig:InflueNonInfleTheshold_4}(b)). If the threshold is changed from \unit[2]{s} to \unit[2.5]{s}, a considerable number of points shown in blue are reclassified as influence points. A mere change of \unit[0.5]{s} in the headway threshold led to a notable difference in the proportion of influence points. This is a drawback of using heuristic-based methods to identify influence points.

\newpage
\subsubsection{Limitations of Hysteresis plots based LF identification}

The core idea of the \cite{wiedemann1974simulation} psycho-physical model revolves around how drivers of faster-moving vehicles respond when approaching slower vehicles. When these drivers reach their perception threshold, they initiate a deceleration process. Nevertheless, due to the inherent challenges in accurately estimating the speed of the lead vehicle, the driver's velocity might temporarily dip below that of the lead vehicle. Consequently, after crossing another threshold, the driver may engage in a minor acceleration phase. This cyclic pattern of acceleration and deceleration occurs iteratively, giving rise to a phenomenon known as hysteresis, which can be attributed to the imperfections in drivers' ability to precisely gauge the exact speeds of the lead vehicles.

\begin{figure}[hbt!]
    \begin{subfigure}{0.47\textwidth}
        \centering
        \includegraphics[width=1\textwidth]{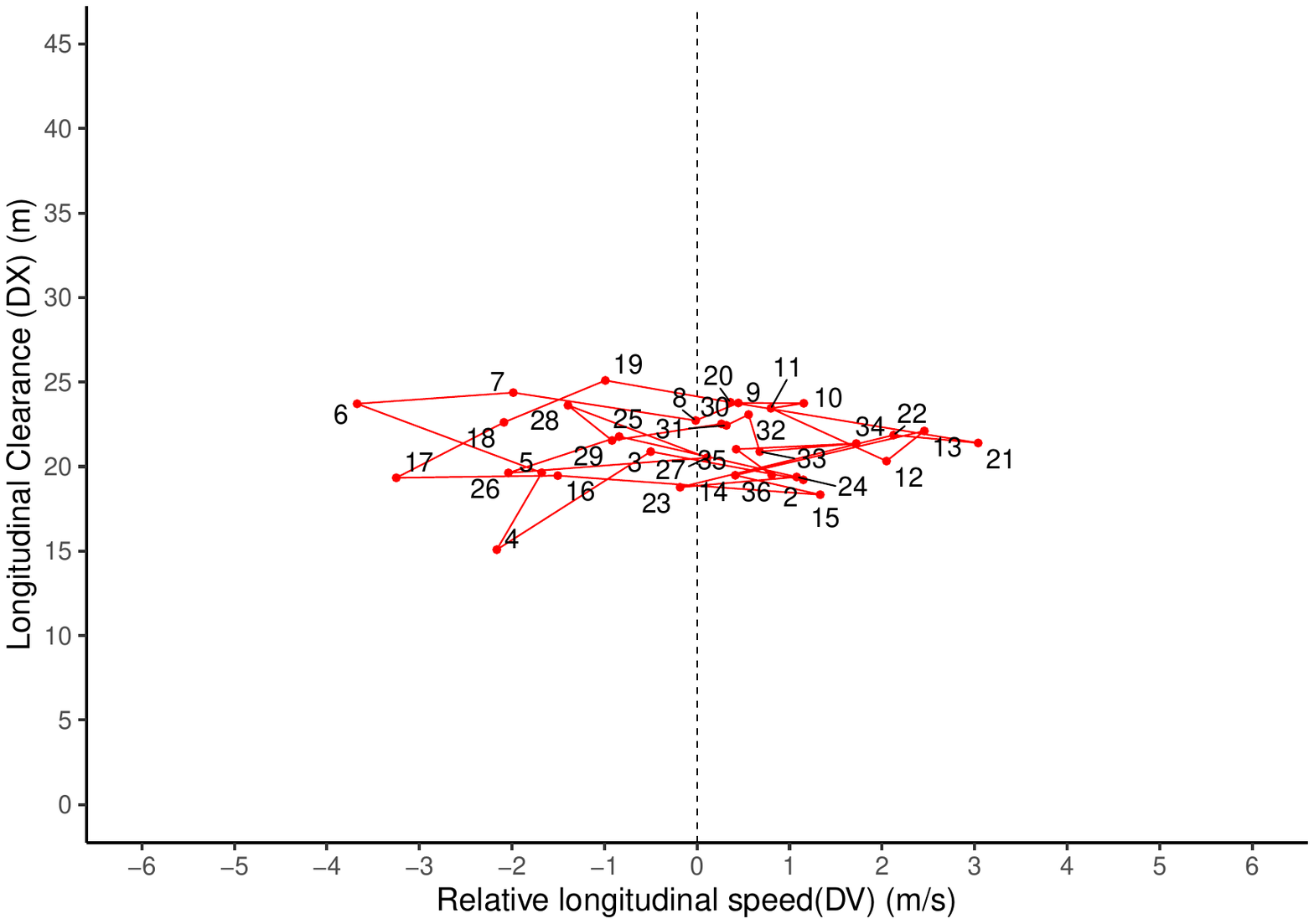}
    \caption{}
    \end{subfigure}
      \begin{subfigure}{0.47\textwidth}
        \centering
        \includegraphics[width=1\textwidth]{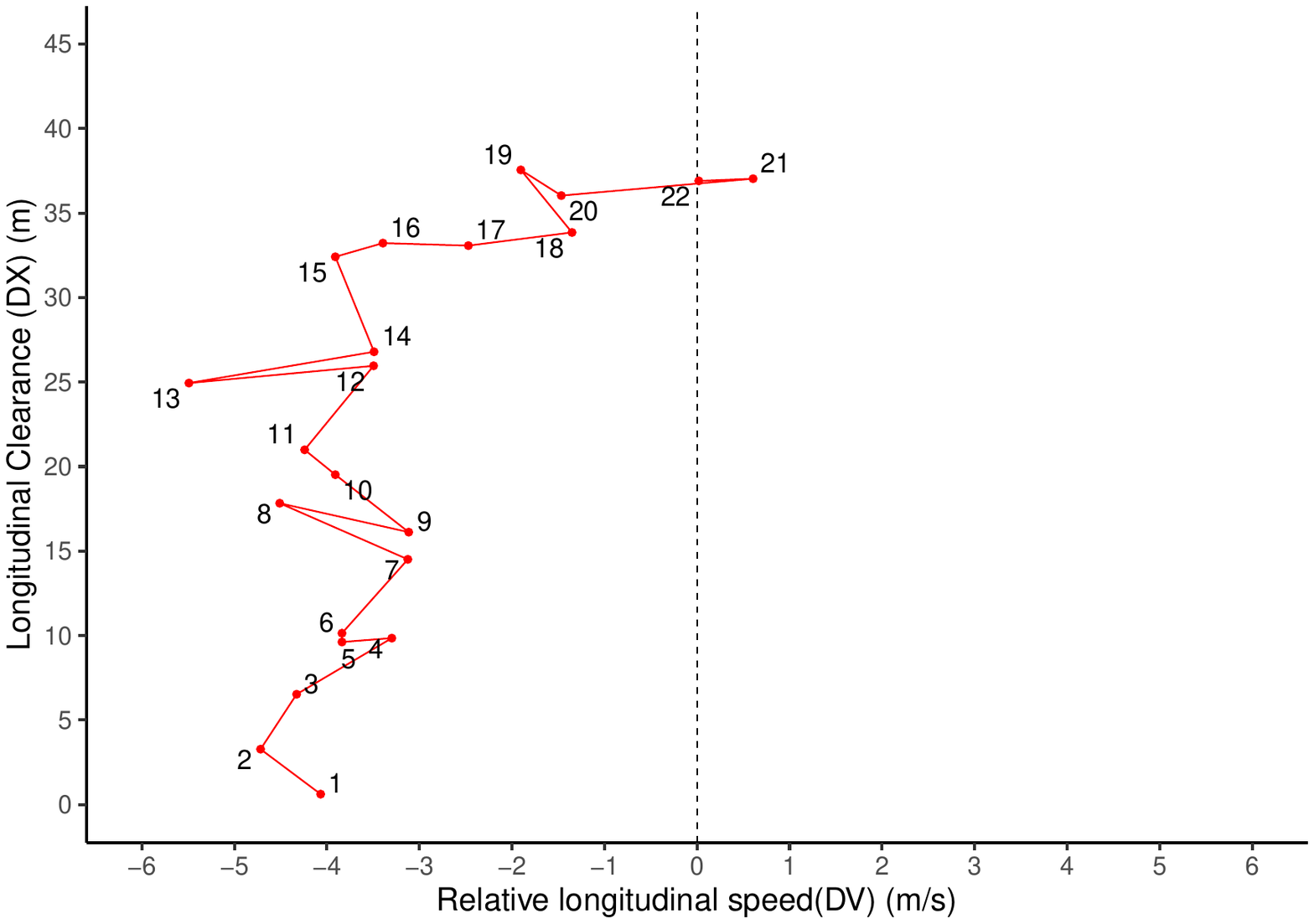}
    \caption{}
    \end{subfigure}
      \begin{subfigure}{0.47\textwidth}
        \centering
        \includegraphics[width=1\textwidth]{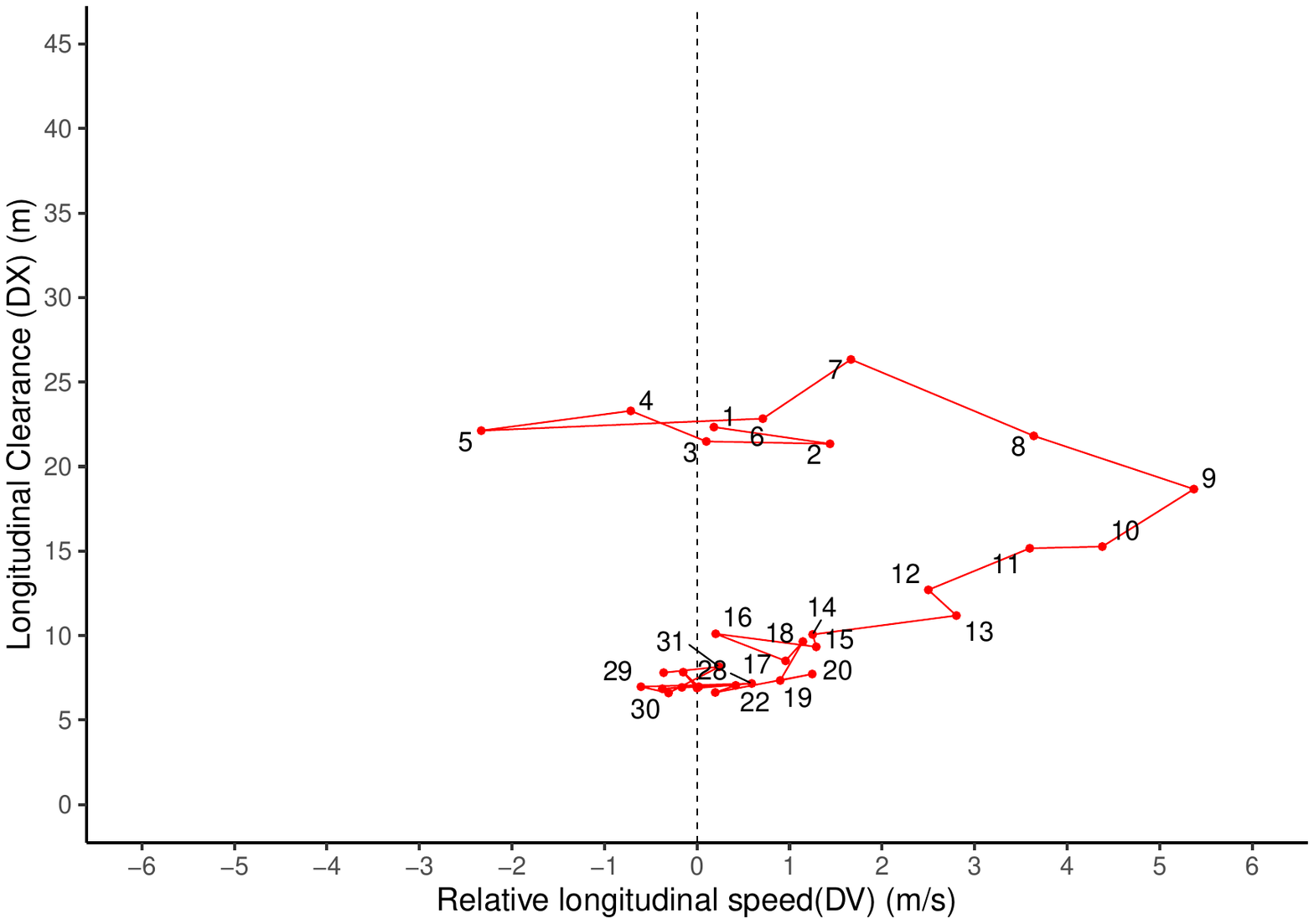}
    \caption{}
    \end{subfigure}
    \begin{subfigure}{0.47\textwidth}
        \centering
        \includegraphics[width=1\textwidth]{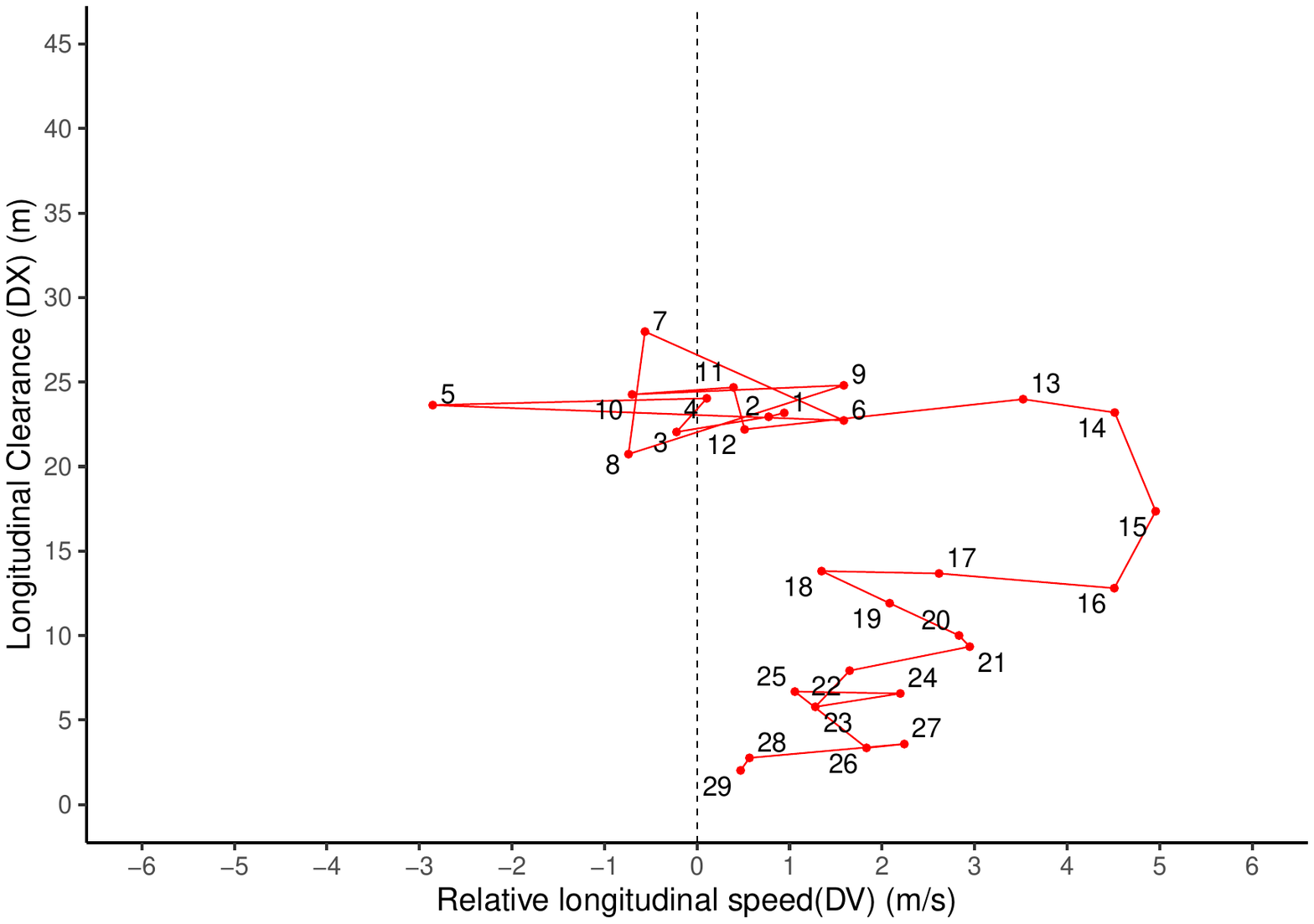}
    \caption{}
    \end{subfigure}
    \caption{Hysteresis plots for LF Identification}\label{Fig:Hysteresis}
\end{figure}

{Figure \ref{Fig:Hysteresis} shows the hysteresis plots of a sample of LF pairs for the adapted method of \cite{raju2021modeling} to identify LF pairs. Numbers associated with points in the figure show time progression. Plot (a) shows clear hysteresis behavior as, for the most part, the longitudinal clearance is almost constant; hence, the follower is trying to maintain the same gap with the leader, and the cyclic pattern is observed at relative speed zero. Thus, the pair can be classified as an LF pair. In plot (b), the leader and follower are clearly not interacting since the leader is significantly faster than the follower, so there is no need for the follower to respond. Plot (c) is interesting since it represents two episodes of the following phase with an approaching phase between points 7 and 14. Similarly, in plot (d), the following phase is only between points 1 to 12, and stimulus-response can be observed only in some parts of the trajectories. Hence, the conclusions could be very subjective by using this method of LF identification.}

\subsection{Performance of LF pairs for calibration of Wiedemann-99 model} {\label{Sec Existing LF}}
In this subsection, LF pairs are identified using the existing methods given in the literature, such as heuristics and hysteresis plots, as explained above. These methods are evaluated by calibration of the Wiedemann-99 model by identified LF pairs. The considered methods are shown below: 

\begin{itemize}
\item M1: All pairs are used.
\item M2: Clearance $<$ \unit[30]{m}, \% lateral overlap $>$ 0\%, continuous following duration $>$ \unit[5]{s}  \cite{anand2019calibration}.
\item M3: Headway $<$ \unit[2]{s} \% lateral overlap $>$ 0\%, continuous following duration $>$ \unit[5]{s} \cite{anand2019calibration}.
\item M4: Clearance $<$ \unit[30]{m}, \% lateral overlap $>$ 50\%, Total following duration $>$ \unit[5]{s}  \cite{anil2022calibrating}.
\item M5: Visual Observation of hysteresis plot \cite{raju2021modeling}.
\end{itemize}

\begin{figure}[h!]
  \centering
  \includegraphics[width=1\textwidth]{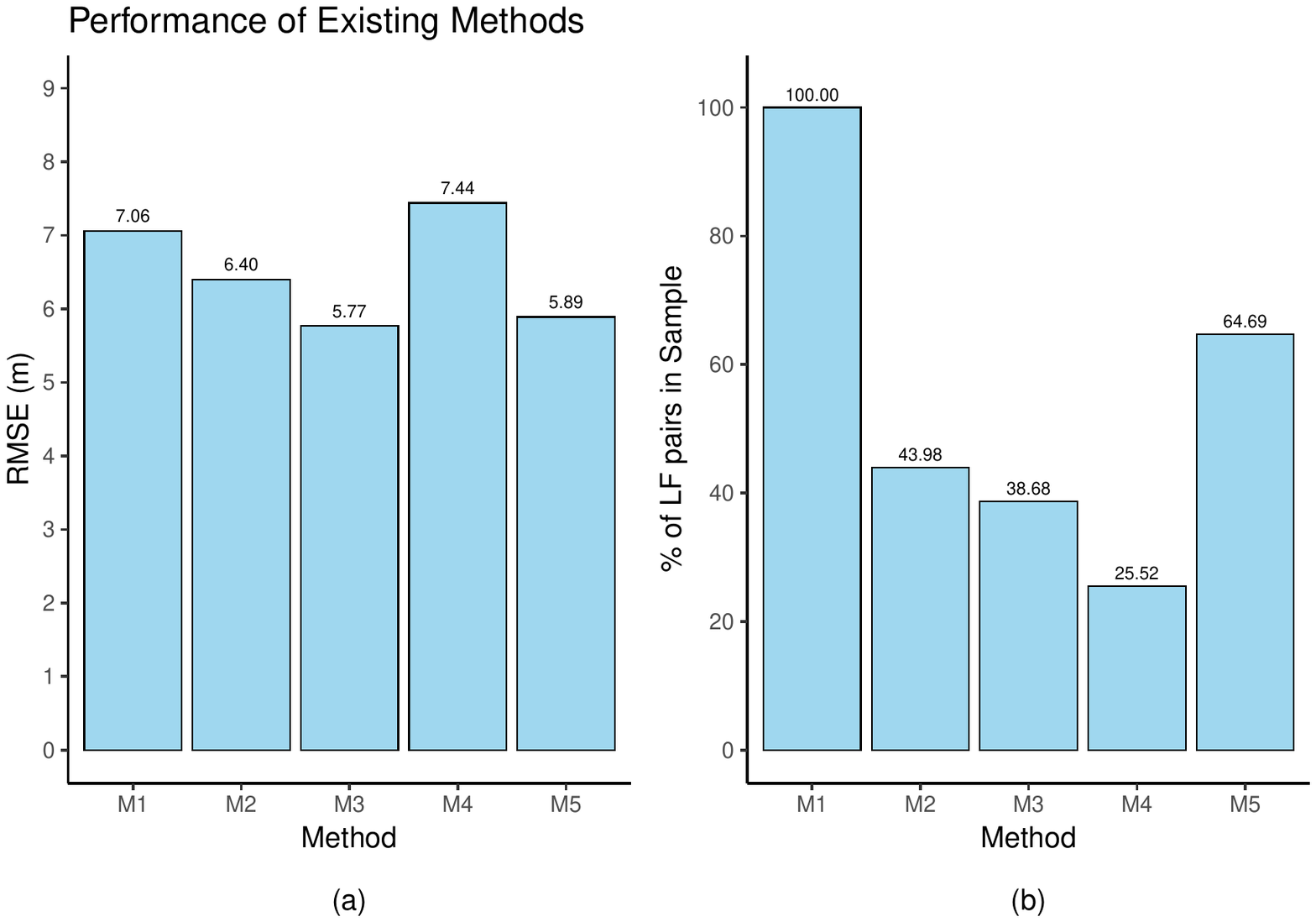}
  \caption{(a) Calibration Performance of existing LF Identification Methods (b)\% of LF Pairs by existing methods}\label{Fig:Existing LF identification methods}
\end{figure}

Figure \ref{Fig:Existing LF identification methods}(a) shows the calibration performance of the Wiedemann-99 model using existing methods for LF identification. The results show that the prediction performance (RMSE) varies across the methods, and it depends on the various arbitrary threshold values, which are subjective. Also, Figure \ref{Fig:Existing LF identification methods} (b) shows the percentage of LF pairs identified by different methods varies substantially, indicating that these methods are not robust. Therefore, there is a need to calculate threshold values tailored to specific traffic conditions which, in turn, influence the calibration results. 
In the following, we propose a method for joint and consistent LF identification and VF model calibration.

\section{Methodology}
This section presents an overview of the proposed optimization-based framework consisting of two tasks:
\begin{enumerate}
    \item LF pair identification process given the set of VF parameters
    \item Calibration of VF parameters is carried out conditional on the set of identified LF pairs from task 1.
\end{enumerate}
\begin{figure}[hbt!]
  \centering
  \includegraphics[width=1\textwidth]{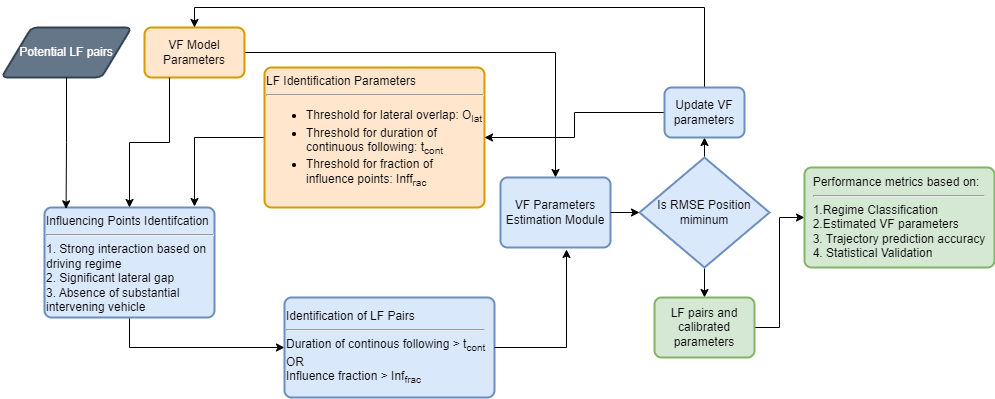}
  \caption{Overview of Proposed Methodology for Joint and Consistent LF Identification and VF parameter estimation}\label{Fig:Methodology}
\end{figure}

Due to their interdependence, the two tasks are carried out iteratively until convergence. Figure \ref{Fig:Methodology} provides a visual summary of the proposed methodology. A brief outline of the two tasks is given below, and a detailed discussion is presented in the following subsections.
The LF identification process invokes three necessary conditions: strong interaction between the subject vehicle and potential leader based on the driving regime, significant lateral overlap between the pair, and the absence of any (substantially) intervening vehicle between them. 

We specify the condition of strong interaction in terms of driving regimes of the Wiedemann-99 psycho-physical model depending on some VF parameters of this model (Sect.~\ref{sec:interaction}). The VF parameters, in turn, are calibrated for identified LF pairs. This mutual dependence between identifying LF pairs and calibrating the VF parameters calls for consistency in estimating the two sets of parameters. 

The remaining criteria, clear gap between the pair (Sect.~\ref{sec:lateral-gap}) and absence of other intervening vehicles (Sect.~\ref{sec:intermediate}), require the calibration of three threshold parameters, namely 1. the gap threshold $c_0$ (negative, if an overlap is required), 
2. the minimum duration $t_{\rm cont}$ of continuous influence as defined by the interaction criterion above for the given potential LF pair, and 3. the minimum fraction $f_{\rm min}$ of (possibly discontinuous) influence between this pair. The lateral gap criterion is applied at each time instance, whereas the remaining two criteria relate to the entire trajectory of the subject vehicle, which is required to have a minimum total duration trajectory {of \unit[5]{s}} (which is not subject to calibration).
A potential LF pair that satisfies the three necessary conditions (either continuously or with intermittent breaks) is deemed to be an actual LF pair. 

In the actual simultaneous calibration process for the VF and LF threshold/identification parameters, we use as initial guesses the standard Wiedemann-99 parameters \citep{W99}{, and $c_0=\unit[0.116]{m}$, $t_{\rm cont}=\unit[5]{s}$, and $f_{\rm min}=\unit[0.36]{}$ as the initial VF parameters.} {Initial guesses are identified by using 1000 randomly generated sets of values, and set for parameters for which RMSE of the position is minimum is considered as an initial guess for optimization.}
With these initial values for both parameter sets, an initial set of LF pairs is identified, and the Wiedemann-99 CF model is estimated by calibration on these LF pairs, minimizing the RMSE of the positions, thus providing the first update of the CF parameters.

Next, LF identification parameters are optimized using a nested optimization approach as described in detail in Sect.~\ref{sec:joint}. These updated VF and LF parameters are fed back into the LF identification process, and the LF identification process is iterated until convergence of both sets of LF pairs and associated VF parameters, i.e., they are now mutually consistent.
Once all LF pairs are determined with calibrated parameters, the results as obtained through different LF identification methods discussed in Sect. \ref{Sec Existing LF} are validated using performance metrics based on Wiedemann-99 regimes classification, estimated VF parameters, trajectory prediction accuracy, and statistical validation via bootstrapping.

\subsection{Determination of influence points}
At a particular instant of time $t$, it is assumed that a subject vehicle is influenced by another vehicle that is immediately ahead of it under the following conditions:

\subsubsection{\label{sec:interaction}Influencing regime criterion} 
The influencing regime criterion assesses whether, for a candidate LF pair, the leader’s motion exerts an influence on the follower’s response as evaluated by the Wiedemann-99 model \citep{W99}. Specifically, the subject vehicle (follower) is considered to be influenced by the leader if it is in either the following, closing, or emergency braking regimes of this model and satisfies the corresponding relative speed and spacing thresholds (see Figure \ref{Fig:Wiedemann} below).

\begin{figure}[hbt!]
  \centering
  \includegraphics[width=1\textwidth]{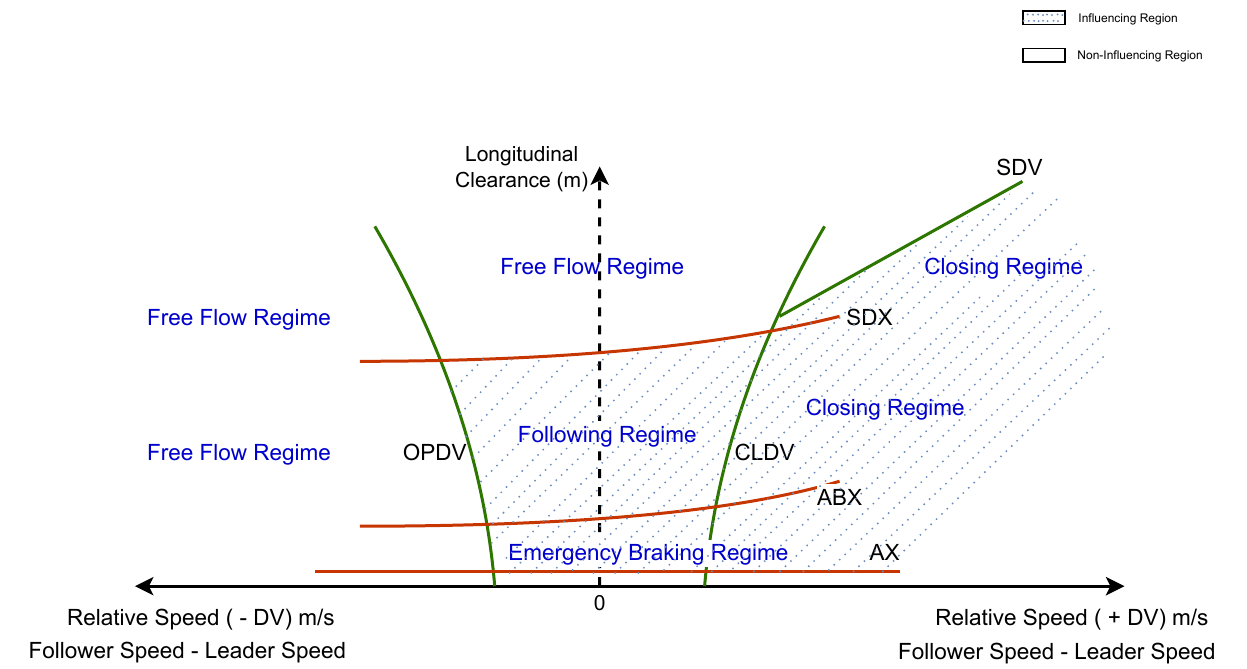}
  \caption{Influencing Region Based on Wiedemann-99 Model \citep{wiedemann1974simulation}}\label{Fig:Wiedemann}
\end{figure}

In order to obtain a numerical criterion, we observe that there is \emph{no} influence if one or more of the following conditions are satisfied: the speed of the vehicle exceeds its free flow speed, the relative spacing exceeds SDX (or the minimum of SDX and SDV if the latter is defined), or the relative speed is smaller than the corresponding OPDV threshold. If none of these three conditions are satisfied, then the point belongs to either following, closing or emergency braking and is thus considered \textbf{an influencing point}. 
It is important to note that these thresholds depend on the vehicle following parameters (CC0, CC1, CC2, CC3, CC4, CC5, and CC6), {as discussed in Appendix A}. Thus, one of the inputs for the influence regime criterion consists of a set of VF parameters necessitating feedback between the VF estimation and LF pair identification procedures.

\subsubsection{\label{sec:lateral-gap}Lateral gap criterion} 
\begin{figure}[hbt!]
  \centering
  \includegraphics[width=0.4\textwidth]{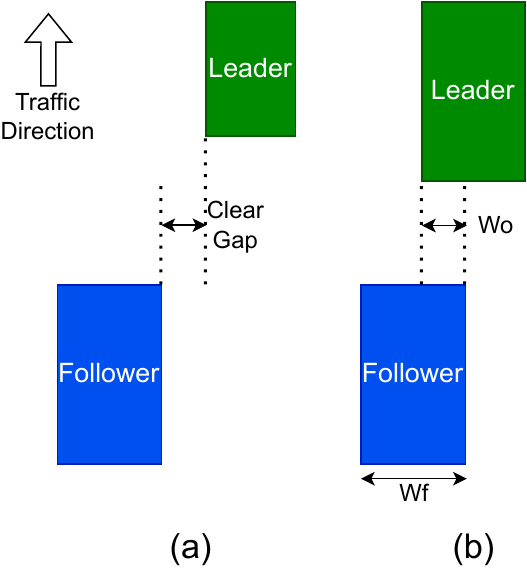}
  
  \caption{Lateral conditions to identify influence points showing ‘lateral clear gap’ and ‘lateral overlap’ criteria.}\label{Fig:Lateral Condition}
\end{figure}

The lateral gap criterion examines if the potential LF pair is in sufficiently close lateral proximity. This condition assumes that the leader can influence the follower when the lateral clear gap is less than the threshold \(c_0\), as shown in Figure \ref{Fig:Lateral Condition}. 

The proposed lateral clear gap criterion is compared with two other lateral criteria in the literature. 

a. Percentage (\%)  lateral overlap \citep{kanagaraj2015trajectory},
\begin{equation}
 W_{\rm o}/W_{\rm f} >O_{\rm lat}.
\end{equation} 
As shown in Figure \ref{Fig:Lateral Condition}, \(W_{\rm o}\) is the overlapping width, \(W_{\rm f}\) is the follower’s width, and \(O_{\rm lat}\) is a calibrated threshold. According to this condition, the leader influences the follower if the lateral overlap relative to the follower's width is greater than the threshold. We emphasize that this criterion is ineffective in the case of a two-wheeler as a leader because it considers width $W_{\rm l}<O_{\rm lat}*W_{\rm f}$ that is exactly centered ahead of the follower as non-interacting.

b. Absolute lateral overlap, \(W_{\rm o}\)$>$\(O_{\rm abs}\)
The percentage overlap condition also depends on the width of the follower, which is not considered by the absolute overlap condition. Both criteria assume a minimal required positive overlap, unlike the proposed clear gap criterion.

\subsubsection{\label{sec:intermediate}No substantial intermediate vehicle effect criterion} 
This criterion checks whether there is any substantially influencing intermediate vehicle between the subject vehicle and its potential leader. If such a vehicle is present, the interaction between the subject vehicle and the potential leader will be significantly weaker. Hence, the associated pair is treated as a non-LF pair. 
For evaluating this criterion, a rectangular intermediate zone (RIZ) is defined as the area enclosed by the horizontal lines representing the rear of the potential leader and front of the subject vehicle and vertical lines representing the left-most and right-most edges of these two vehicles. Notice that this criterion is only evaluated, and an RIZ is only defined if a candidate LF pair interacts according to the Wiedemann regimes and satisfies the lateral gap criterion because all criteria for a valid LF pair must be satisfied simultaneously.

\begin{figure}[hbt!]
  \centering
  \includegraphics[width=1\textwidth]{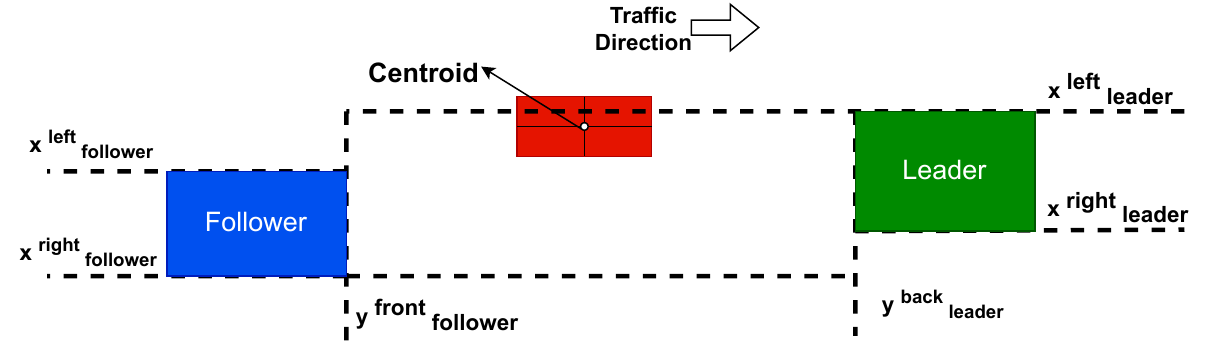}
  \caption{Rectangular intermediate zone (RIZ) using the proposed method.}\label{Fig:RIZ}
\end{figure}

The rectangular intermediate zone is defined as shown in Figure \ref{Fig:RIZ} by dashed lines, where,
\begin{itemize}
    \item Left boundary = \(min(x_{\rm leader}^{\rm left},x_{\rm follower}^{\rm left})\);
\item Right boundary = \(max(x_{\rm leader}^{\rm right},x_{\rm follower}^{\rm right})\) ;
\item Back boundary = \(y_{\rm follower}^{\rm front}\);
\item Front boundary = \(y_{\rm leader}^{\rm back}\).
 
\end{itemize}

Based on the presence or absence of an intermediate vehicle in this zone (RIZ), as well as the extent of the presence of the intermediate vehicle (corner, edge, centroid presence) and overlap with potential leader and follower, six possible cases are delineated as shown in Figure \ref{Fig:In between}.  
During the analysis of all leader-follower (LF) pairs, each data point is scrutinized to determine the influence of other surrounding vehicles on the considered LF pair. It is based on the factors such as vehicle sizes, positions, and lateral overlaps. The various scenarios of intermediate vehicle effects are shown in Figure \ref{Fig:In between} and described below:

\begin{figure}[hbt!]
  \centering
  \includegraphics[width=1\textwidth]{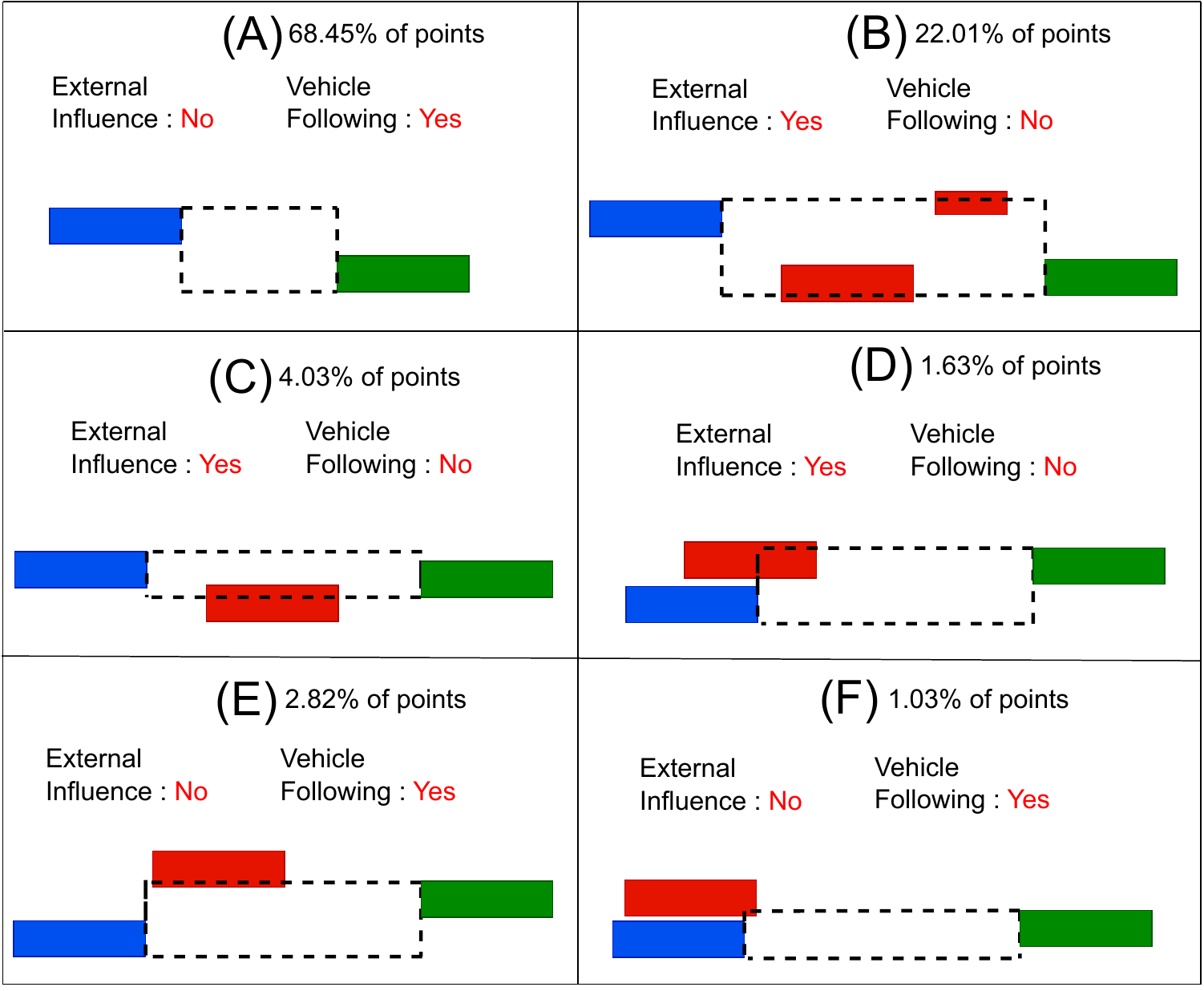}
  \caption{Different categories of identification of external vehicle’s influence on assumed leader and follower.}\label{Fig:In between}
\end{figure}

\begin{enumerate}[label=(\Alph*)]

\item \textbf{No part of a vehicle is present in RIZ}: This is a clear case that the intermediate vehicle plays no role between the potential LF pair. This case indicates the absence of external influence and, consequently, the likelihood of vehicle-following between leader and follower provided other conditions such as driving regime and lateral overlap are satisfied. In the empirical data, 39829 {time} instances of this category were observed, where no other vehicle is partially or fully detected within the rectangular boundary.

\item \textbf{Centroid (as shown in Figure \ref{Fig:RIZ}) of at least one vehicle is present in RIZ}: The presence of centroid indicates a strong external influence of this intervening vehicle, which disrupts vehicle-following between the subject vehicle and its potential leader. In this dataset, 12805 time instances of this category are observed. During these instances, due to this intermediate vehicle, the interaction between the subject vehicle and the potential leader is assumed to be non-following.  

\item \textbf{Intermediate vehicle’s edges(s) is (are) detected in the RIZ ahead of the follower and has lateral overlap with it}: Although only the corner of the intermediate vehicle is present, the considerable lateral overlap with leader suggests a notable influence of the intervening vehicle. In the 2341 time instances of this category, it is assumed that due to the effect of the vehicle in-between, there is no following between the potential leader and the subject vehicle. {(If a time instance satisfies both (B) and (C), it is counted for Criterion (B). Generally, a time instance is counted for the first criterion that applies.)}

\item \textbf{Intermediate Vehicle’s corner is detected in RIZ whose back edge is behind the follower’s front edge, and has greater lateral overlap with the leader than the lateral overlap between potential leader and follower}: This is another instance where despite the small presence of an intermediate vehicle in the RIZ, it has a significant effect on disrupting vehicle following behavior due to the considerable overlap with the leader. Hence, in the 974 time instances of this category, the intermediate vehicle has a stronger effect than {(F)} and is assumed to result in non-following.

 \item \textbf{Intermediate vehicle’s corner(s) is (are) detected in the RIZ ahead of the follower but without any lateral overlap}: In this case, there is the marginal presence of intermediate vehicle, but unlike case C it is much weaker due to the lack of overlap with subject vehicle. Among the 1643 instances of this category in the sample data, it is assumed that the intermediate vehicle is staggered away from the LF pair and does not interfere with the vehicle-following process.

\item \textbf{Intermediate Vehicle’s corner is detected in RIZ whose back edge is behind the follower’s front edge and has smaller lateral overlap with the leader than the lateral overlap between potential leader and follower}: Due to the lower marginal presence of an intermediate vehicle and lesser overlap with the leader, in the 599 instances of this category, the leader-follower interaction is not affected much by the intermediate vehicle. Hence vehicle-following behavior is plausible for the LF pair.
\end{enumerate}

For each case, the intermediate vehicle's potential presence or absence of influence on the potential LF pair is also analyzed.  From Figure \ref{Fig:In between}, it is clear that cases B, C, and D correspond to the potential influence of the intermediate vehicle; in the case of C and D, the intermediate vehicle's centroid doesn't need to be present in RIZ., whereas A, E, and F represent the relatively little impact of the intermediate vehicle. Thus, A, E, and F could represent influencing points where the subject vehicle in blue is directly influenced by the leader ahead in green, whereas, in the remaining cases, the subject vehicle will not be considered to be influenced by the leader due to the intervening vehicle effect. 
The instant of time, when all these conditions are satisfied for a specific pair of follower and leading vehicles, is considered an influence point for this pair. For a given potential LF pair, the subject vehicle is under a given lead vehicle's influence at some time instants but not at others.

\subsection{Determination of LF pairs from influence points}
The presence of an influence point is suggestive but not confirmatory of a following behavior between the leading and following vehicles. For instance, the two vehicles may come into mutual influence even during lane change or lateral movements, which may be momentary or temporary. Thus, influence points are necessary but not sufficient to determine whether a given pair of vehicles is indeed a leader-follower pair. 
To resolve this ambiguity, it is postulated that for a subject vehicle that is following a lead vehicle, either the duration of continuous influence and/or the fraction of total influence points will be substantial. Accordingly, a given pair of vehicles is said to be in leader-follower mode if one (or both) of the following conditions is satisfied:
\begin{enumerate}
    \item 	Duration of continuous influence points ($t$) exceeds a minimum threshold (\(t_{\rm cont}\)) or
    \item {The percentage of influence points in the subject vehicle’s trajectory is greater than a given threshold $f_{\rm min}$.} 
 
\end{enumerate}
The first condition implies that if the continuous duration when the subject vehicle is likely to be influenced by the vehicle ahead is substantial (say more than 1 minute), this is consistent with the subject vehicle following the leader. The second condition implies that even if the duration of influence is not continuous, but the subject vehicle was in close and direct proximity of the vehicle ahead for a substantial fraction of its trip (say 50\%), the subject vehicle may be in the following mode.

\subsection{VF Parameter Estimation}

\begin{algorithm}

    \caption{Calibration Methodology, adopted from \cite{anil2022calibrating}}\label{alg:calibration}
    \begin{algorithmic}[1]
    \State \textbf{Initialisation:}
    \State Initial Wiedemann-99 (CC) parameters\\
    \For{$follower (i) = 1,2,....,N$} \Comment{ $N$ is number of followers}
        \State \(a^{\rm sim}_{i}(t=1) = a^{\rm obs}_{i}(t=1)\)
         \State \(v^{\rm sim}_{i}(t=1) = v^{\rm obs}_{i}(t=1)\)
            \State \(x^{\rm sim}_{i}(t=1) = x^{\rm obs}_{i}(t=1)\)\\
   
        \For { $iteration (t)=1, 2, \ldots, (T_i-1)$} \Comment{$t=1$ is the begin of the leader-follower episode and $T_i$ is the duration of leader-follower episode $i$ }
        
            \begin{enumerate}
            \item Calculation of Wiedemann-99 thresholds AX, ABX, SDX, CLDV, OPDV, and SDV 
            \item Identification of Regimes (Closing, Following, Free-flow, Emergency Braking, and Opening) 
            \item Prediction of acceleration \(a^{\rm sim}_{i}(t+\tau) \) 
            \item Prediction of Speed \(v^{\rm sim}_{i}(t+\tau)\) and Position \(x^{\rm sim}_{i} (t+\tau)\) using numerical integration method 
            \end{enumerate}
        \EndFor
    \State Predicted trajectory for follower $i$
    \State  \(RMSE(x)_i=\sqrt{\frac{1}{T}_{i}\Sigma_{t=1}^{{T}_{i}}{[x^{\rm obs}_{i}(t)-x^{\rm sim}_{i}(t)]^2}}\)   

   \EndFor
   \State \(Z (CC0,...,CC7) = \frac{\Sigma_{i=1}^{N}RMSE(x)_i}{N}\)
   \State \(Minimize\) \(Z\) to get optimized Wiedemann-99 (CC) Parameters

    \end{algorithmic}
\end{algorithm}


Once the LF pairs are identified by the above procedure (Section 4.2) for an assumed initial set of VF parameters, the VF parameter estimation is performed by calibration, as shown in Algorithm 1 and briefly outlined below. The leader trajectory is assumed to be fixed. Based on initial VF parameters, regimes are identified for each subject vehicle (follower) and each time $t$. For the corresponding regime, the modified Wiedemann-99 model equations (as modified in \cite{anil2022calibrating}) are used to compute the acceleration response of the subject vehicle at time $t$+ $\tau$ where $\tau$ represents the timestep. Next, numerical integration is used to determine the speed at the next time interval $t$+ $\tau$, which is then integrated to compute the position at the next time step. For this updated time step, the longitudinal gap and speed difference with the leader are estimated, and the time step is incremented. This process is continued until all points of the subject vehicle and all subject vehicles are processed. The resultant output is the predicted trajectory (position, speed, and acceleration profile) of the subject vehicle. The RMSE in position between the predicted and actual trajectory of the subject vehicle is computed and minimized to obtain the desired VF parameters. \cite{punzo2021calibration} concluded that calibration on spacing is {to be favored over speed} as minimization of spacing errors is equivalent to simultaneous minimization and optimal time allocation of speed errors. Algorithm 1 gives an overview of the VF calibration procedure. 

\subsection{\label{sec:joint}Joint and Consistent Estimation Procedure}

The proposed methodology aims to determine LF pairs and calibrate the Wiedemann-99 car-following model jointly and consistently. 

Overall, the combined LF identification and vehicle following model have a total of 10 parameters, 7 out of 10 CC parameters for Wiedemann-99 models, {and 3 for identification of LF pairs. Notice that the parameters CC0, CC6, and CC9 are not considered for calibration. According to \cite{anil2022calibrating}, CC0=\unit[0.65]{m} is a constant for a given vehicle type (here, cars), CC6 is assumed as the default value, and CC9 is not used in the modified acceleration equations.}
The LF identification thresholds include parameters on lateral clearance (or other lateral criteria), duration of continuous following, and fraction of influential points. 

The joint and consistent estimation is accomplished by creating feedback between the LF identification procedure, which takes VF parameters as input and determines driving regimes, which in turn form the basis of classification into influencing and non-influencing points. The resultant (tentative) LF pairs are then fed into the VF model calibration procedure, and the VF parameters are obtained. The updated VF parameters are then used in the next iteration of LF identification, and the process continues until the convergence of both LF identification and VF model parameters, which are mutually interdependent. One key advantage of this procedure is that it provides an optimization method to determine important thresholds for LF pair identification in lieu of trial and error methods currently used in literature.
The objective function used in the above calibration is the deviation between observed and computed longitudinal positions. Thus, calibration is formulated as an optimization problem to determine the best set of model parameter values that minimizes this RMS error. The objective function is given by

\begin{equation}
\label{objZ}
    {Z (\vec{\beta}_{\rm w}, \vec{\beta}_{\rm LF})} = \frac{\Sigma_{i=1}^{N}\sqrt{\frac{1}{T_i}\Sigma_{t=1}^{T_i} [x^{\rm obs}(i,t)-x^{\rm sim}(i,t,\vec{\beta}_{\rm w})]^2}}{N}
   {\ \stackrel{!}{=}\textrm{min}}
\end{equation}
where, Wiedemann parameters  $\vec{\beta}_{\rm w}=(CC0, ..., CC8)'$, LF parameters $\vec{\beta}_{\rm LF}=(\tau_{\rm cont}, f_{\rm min}, c_0)'$, $T_i=T_i(\vec{\beta}_{\rm wr},\vec{\beta}_{\rm LF})$, N=N\((\vec{\beta}_{\rm wr},\vec{\beta}_{\rm LF})\) is the number of LF pairs that are used for model calibration. \(x^{\rm obs}(i,t)-x^{\rm sim}(i,t,\vec{\beta}_{\rm w})\) is the difference between observed and simulated longitudinal coordinates of the follower for pair $i$, \(x^{\rm sim}(i,t)\) is obtained from the numerical integration of velocity which is obtained from the numerical integration of acceleration. Notice that, because of initially setting the simulated position to the data, the objective function~(\ref{objZ}) is equivalent to minimizing the RMS of the gap deviations.

\begin{figure}[hbt!]
  \centering
  \includegraphics[width=1\textwidth]{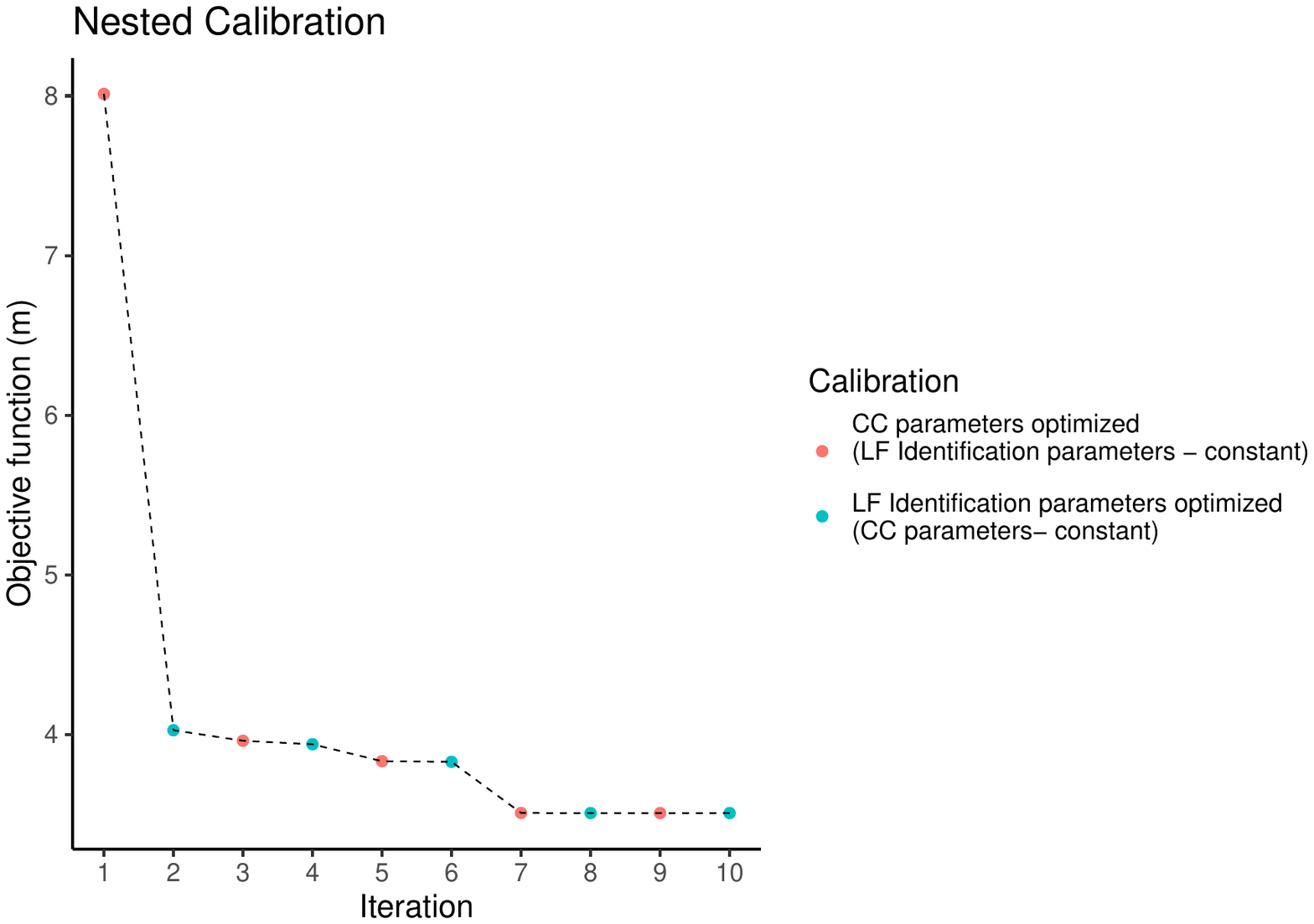}
  \caption{Convergence of Objective Function with the Nested Calibration}\label{Fig:nested}
\end{figure}

A nested calibration approach is used for this calibration, wherein, initially, the LF parameters $\vec{\beta}_{\rm LF}$ are fixed, and the Wiedemann parameters $\vec{\beta}_{\rm w}$ are estimated. In the next iteration, $\vec{\beta}_{\rm w}$ is fixed, and $\vec{\beta}_{\rm LF}$ is estimated, and so on. The process is repeated until both CC and LF (indicated as LF identification parameters in Figure \ref{Fig:nested}) parameters converge. Figure \ref{Fig:nested} shows the convergence pattern of the objective function. The nested optimization process is more stable than a direct simultaneous estimation of all the 10 parameters of ($\vec{\beta}_{\rm LF}, \vec{\beta}_{\rm w})$ because the number of LF pairs depends on \((\vec{\beta}_{\rm LF})\) resulting in discontinuities in the objective function. Using the nested process restricts the discontinuities to the calibration of the three LF parameters increasing the stability and performance of the whole optimization process.

To summarize, a new joint and consistent framework is proposed to estimate LF identification and vehicle following parameters simultaneously. This procedure takes as inputs trajectory data and initial VF and LF starting values. By optimizing the RMSE of position, the procedure produces the following outputs: RMSE of position, number, and set of leader-follower pairs, converged VF parameters, and LF identification thresholds. The performance of the proposed procedure is evaluated with respect to other LF identification methods from the literature in the following results section. 

\section{Results and Findings}
This section presents a comparison between existing LF identification methods and the proposed approaches. The comparison is conducted based on several criteria, including \(RMSE\) of position, the number of LF pairs, regime distributions of LF and non-LF pairs, and performance on the validation set. Additionally, a statistical comparison is made between different VF parameters using bootstrapping.\
Table 2 provides a detailed description of the various LF identification methods (existing and proposed) utilized in this study, calibrated (non-calibrated for some methods) values of LF identification parameters, additional thresholds required by each method for identifying LF pairs, Goodness of Fit \((GOF)\) which is \(RMSE\) of position for each method and total number of LF pairs identified by each method {(cf. Table~\ref{tab:methods}).}
\begin{enumerate}[label=(\alph*)]
    \item \textbf{M1-M5} represent existing methods without modifying heuristic thresholds, as explained in Section 3.3.
    \item \textbf{M6-M7} existing methods (M2 and M3 respectively) have been adapted with the proposed approach of joint and consistent calibration. In these cases, thresholds are calibrated rather than relying on heuristic values.
    \item \textbf{M8} is {our} proposed method that takes into account and calibrates the clear gap threshold {according to Algorithm~\ref{alg:calibration}}
    \textbf{M9}  is {Method M8} without calibration of LF identification parameters.
    \item  \textbf{M10-M11} are methods from the literature that involve calibrating LF identification parameters. \textbf{M10} is based on visual angle.\citep{yousif2011close}.
    \textbf{M11} uses Delaunay triangulation similar to \citep{nagahama2021detection} to identify intermediate vehicle.
    \item \textbf{M12-M13} are modified versions of the proposed methods (M8) considering different heterogeneous features such as Absolute Lateral Overlap and \% lateral overlap instead of the lateral clear gap as proposed.

\end{enumerate}
   
\newpage
\begin{table}[!ht]
     \resizebox{\textwidth}{!}
    {
    \begin{threeparttable}[b]
    \centering
    \caption{LF identification parameters for different methods}\label{tab:methods}
    \fontsize{10pt}{12pt}\selectfont
    \renewcommand{\arraystretch}{1.2}
    \renewcommand{\thempfootnote}{\arabic{mpfootnote}} 
   
    \begin{tabular}{p{1cm}p{4cm}p{1cm}p{1cm}p{1cm}p{2.5cm}p{1cm}p{1cm}}
    \hline
    \textbf{Method} & \textbf{Description} & \textbf{O\_lat} & \(t_{\rm cont}\) (Sec) & \textbf{$f_{\rm min}$} & \textbf{Additional Thresholds} & \textbf{GoF} (m) & \textbf{Number of LF pairs} \\
    \hline
    M1 & All Pairs & -- & -- & -- & -- &7.06 &623\\
    
    M2 & Longitudinal Clearance Heuristic (Neither Joint nor Consistent) & 0.0\%\tnote{*} & 5\tnote{*} & -- & Longitudinal Clearance = 30 m\tnote{*} &6.40 &274\\
    
    M3 & Headway Heuristic (Neither Joint nor Consistent) & 0.0\%\tnote{*} & 5\tnote{*} & -- & Headway = \unit[2]{s}\tnote{*} &5.77 &241\\
    
    M4 & Influence Area Heuristic (Neither Joint nor Consistent) & 50.0\%\tnote{*} & -- & -- & Longitudinal Clearance = 30 m\tnote{*}, Total Duration of Following = \unit[5]{s}\tnote{*} &7.44 &159 \\
    
    M5 & Hysteresis Plots-based LF Identification by \cite{raju2021modeling} (Neither Joint nor Consistent) & -- & -- & -- & -- &5.89 &403\\

    M6 & Longitudinal Clearance Heuristic (Joint and Consistent) & 33.0\% & 3.5 & 0.66 & Longitudinal Clearance = 30 m\tnote{*} &6.25 &199 \\

    M7 & Headway Heuristic (Joint and Consistent) & 63.0\% & 6.5 & 0.60 & Headway = \unit[2]{s}\tnote{*} &3.65 &44\\
    
    M8 & Lateral Clear Gap (Joint and Consistent) & -- & 7 & 0.54 & {Lateral }Clear Gap Threshold: 0.157m &3.51 &100 \\
    
    M9 & Lateral Clear Gap (Neither Joint nor Consistent) & -- & 5\tnote{*} & 0.35\tnote{*} & Lateral Clear Gap Threshold: 0.116m &4.03 &143\\
    
    M10 & Visual Angle (Joint and Consistent) & -1.0\%\tnote{\#} & 7 & 0.65 & Visual Angle Threshold = 0.0029 &6.40 &253\\
    
    M11 & Voronoi Triangulation (Joint and Consistent) & -- & 6.5 & 0.57 & Clear Gap Threshold = 0.3m\tnote{*} &3.62 &116\\
    
    M12 & Absolute Lateral Overlap (Joint and Consistent) & -- & 5 & 0.45 & 0.007 m &3.83 &107\\
    
    M13 & Lateral \% Overlap (Joint and Consistent) & -- & 4.5 & 0.55 & \% Lateral Overlap Threshold = 0.06\% &3.89 &108\\
    \hline
    \end{tabular}
   
    \begin{tablenotes}
       \item [--] Threshold is not applicable
       \item [*] Uncalibrated or initial values
       \item [\#] -1\% indicates that influence would be there even for no overlap
       \item  Calibrated CC parameters of M8: CC0: \unit[0.65]{m}; CC1: \unit[0.30]{s}; CC2: \unit[7.00]{m}; CC3: \unit[-3.90]{s}; CC4: \unit[-2.34]{m/s}; CC5: \unit[2.41]{m/s}; CC6: \unit[11.44]{1/ms}; CC7: \unit[0.61]{m/$s^2$}; CC8: \unit[3.09]{m/$s^2$}
       
     \end{tablenotes}
    \end{threeparttable}
    }
\end{table}

\subsection{Benefits of Joint and Consistent Calibration}
\begin{figure}[hbt!]

  \centering
  \includegraphics[width=1\textwidth]{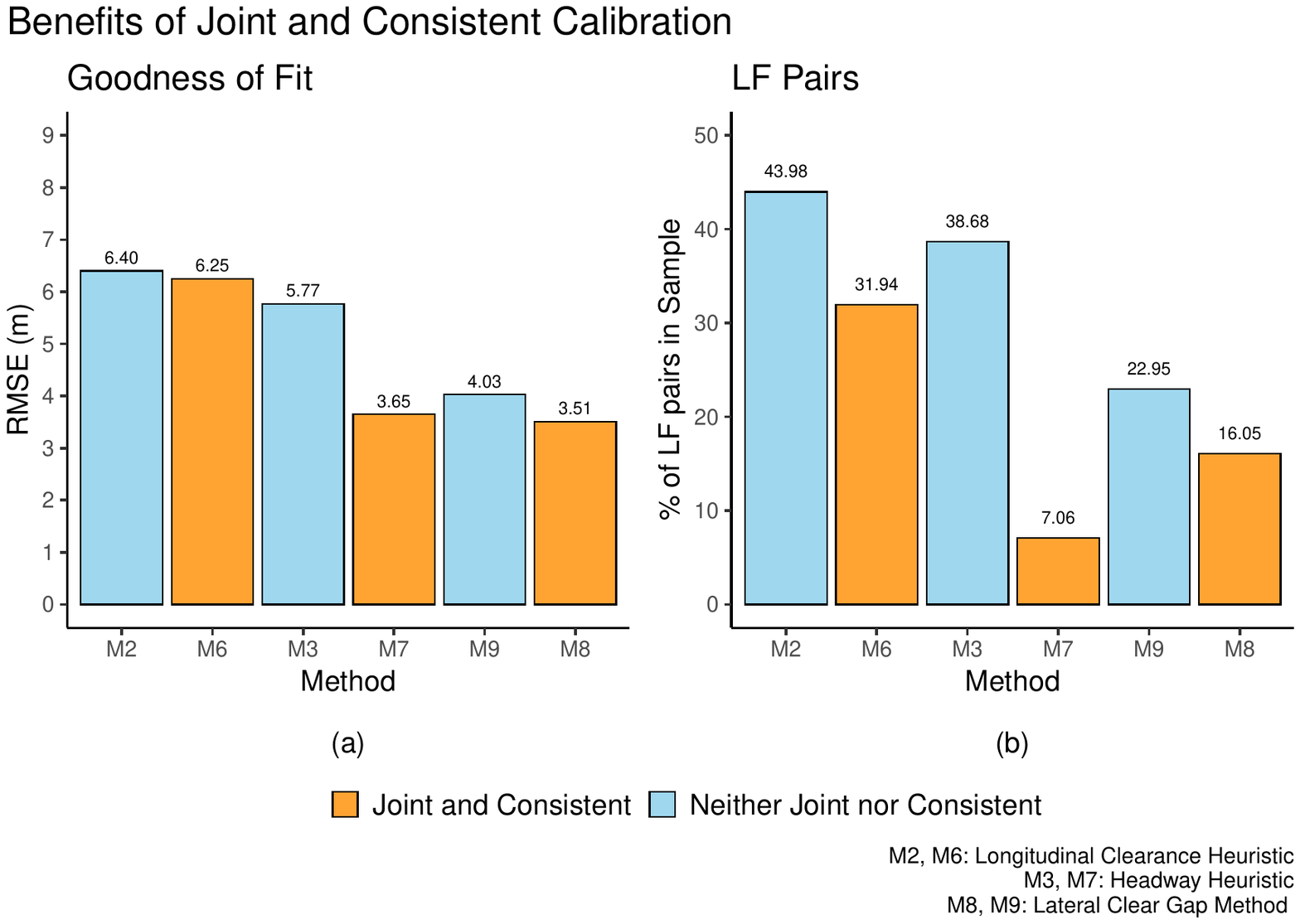}
  \caption{Benefits of joint and consistent calibration}\label{Fig:benefitofjoint}
  
\end{figure}

Figure \ref{Fig:benefitofjoint}(a) shows the benefit of joint and consistent calibration. For the longitudinal clearance heuristic (M2), headway heuristic (M3), and lateral clear gap model (M9), we can see an improvement in RMSE for the corresponding  ‘joint and consistent’ calibration methods M6, M7, and M8 respectively. As shown in Figure \ref{Fig:benefitofjoint}(b), the number of LF pairs identified by the joint and consistent calibration are lower than that of the corresponding uncalibrated models. For M7, the number of LF pairs drops down considerably because the calibrated value of the minimum overlap \(O_{lat}\) for joint and consistent calibration is 63\%, which implies a strict following condition. Based on the performance metric, the proposed model M8 with joint and consistent calibration shows the best results with the lowest GoF measure at a reasonable number of LF pairs.

\subsection{Advantage of psychophysical regime based LF identification}%

\begin{figure}[hbt!]
  \centering
  \includegraphics[width=1\textwidth]{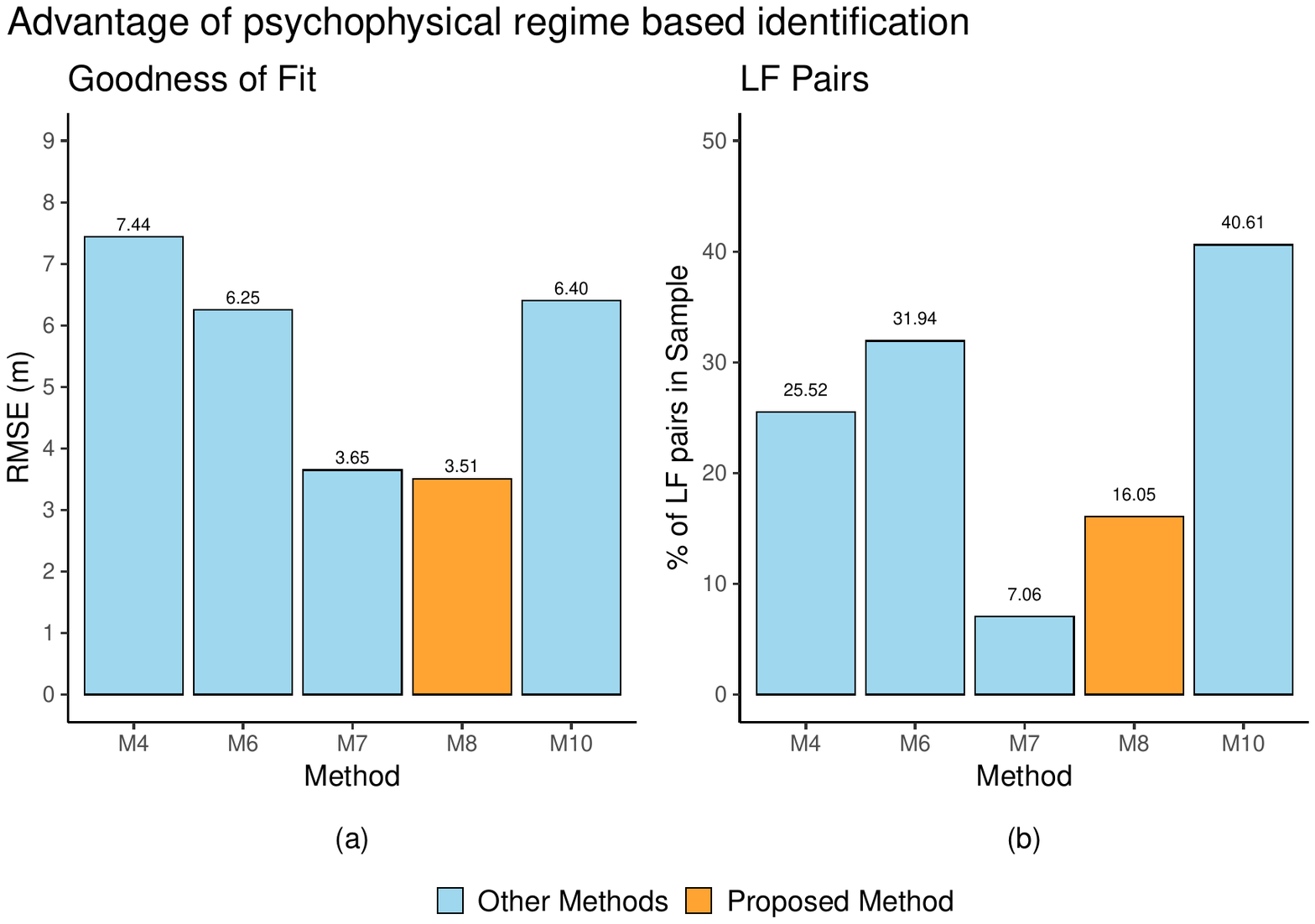}
  \caption{Advantage of psychophysical regime based LF identification.}\label{Fig:advantage of psychophysical}
\end{figure}

Figure \ref{Fig:advantage of psychophysical}(a) points out the advantage of psychophysical regime-based LF identification. Only the proposed model (M8) uses a psychophysical concept for LF identification. For more details of the visual angle model (M10), the reader is referred to \cite{yousif2011close}. 
We calibrate the angular velocity value for the visual angle model in the suggested range. In terms of RMSE, the proposed model performs the best. The performance of the joint and consistent headway heuristic (M7) is closer to that of the proposed model. In Figure \ref{Fig:advantage of psychophysical}(b), a wide variation can be observed in the number of LF pairs identified by each of the models. This shows the difference in selectivity for LF pairs across methods.

\subsection{Regime Distribution of LF and Non-LF pairs}

\begin{figure}[!hbt]

    \begin{subfigure}{1\textwidth}
      \centering
      \includegraphics[width=1\textwidth]{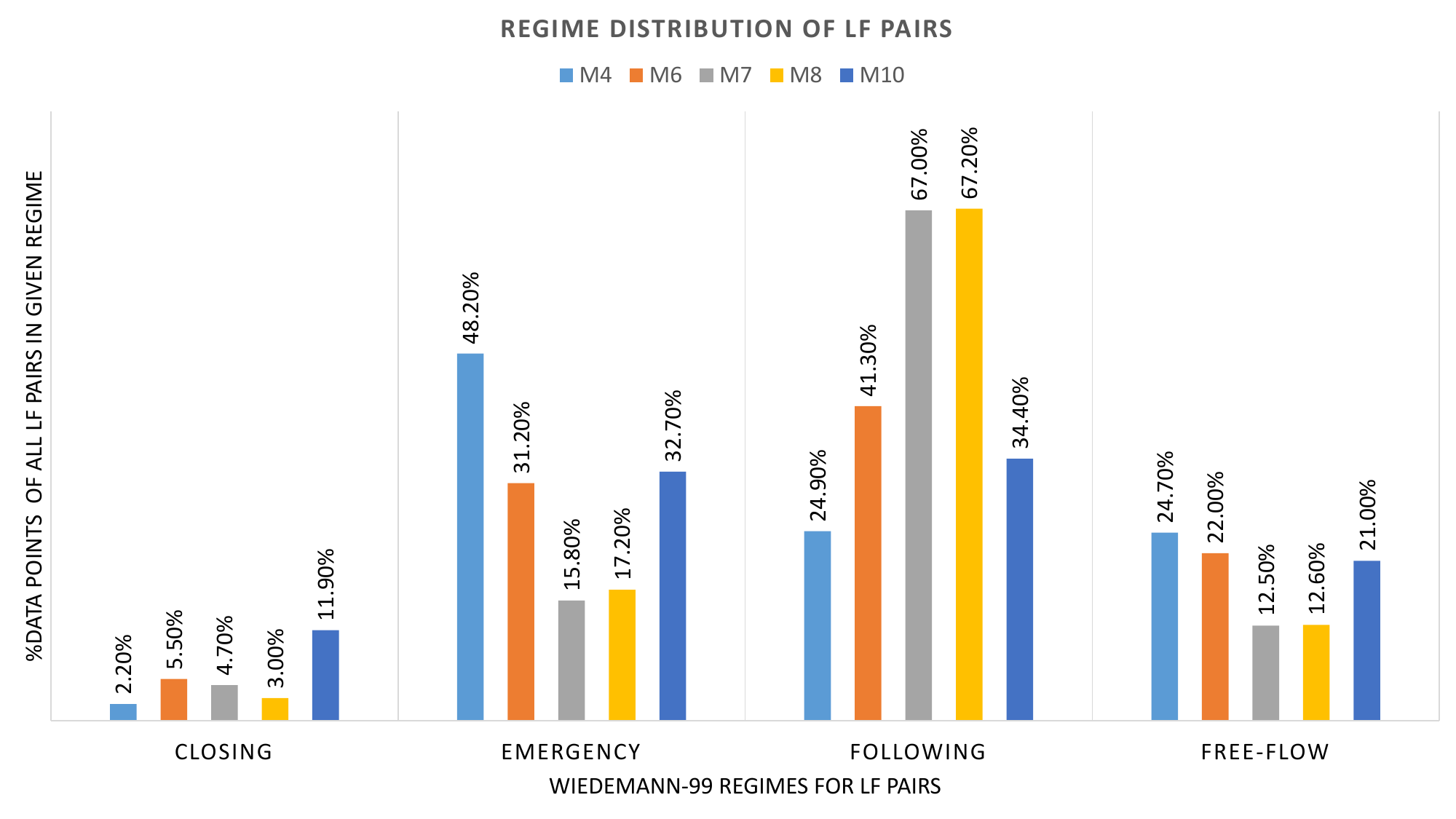} 
      \caption{}
      \end{subfigure}
      \begin{subfigure}{1\textwidth}
      \includegraphics[width=1\textwidth]{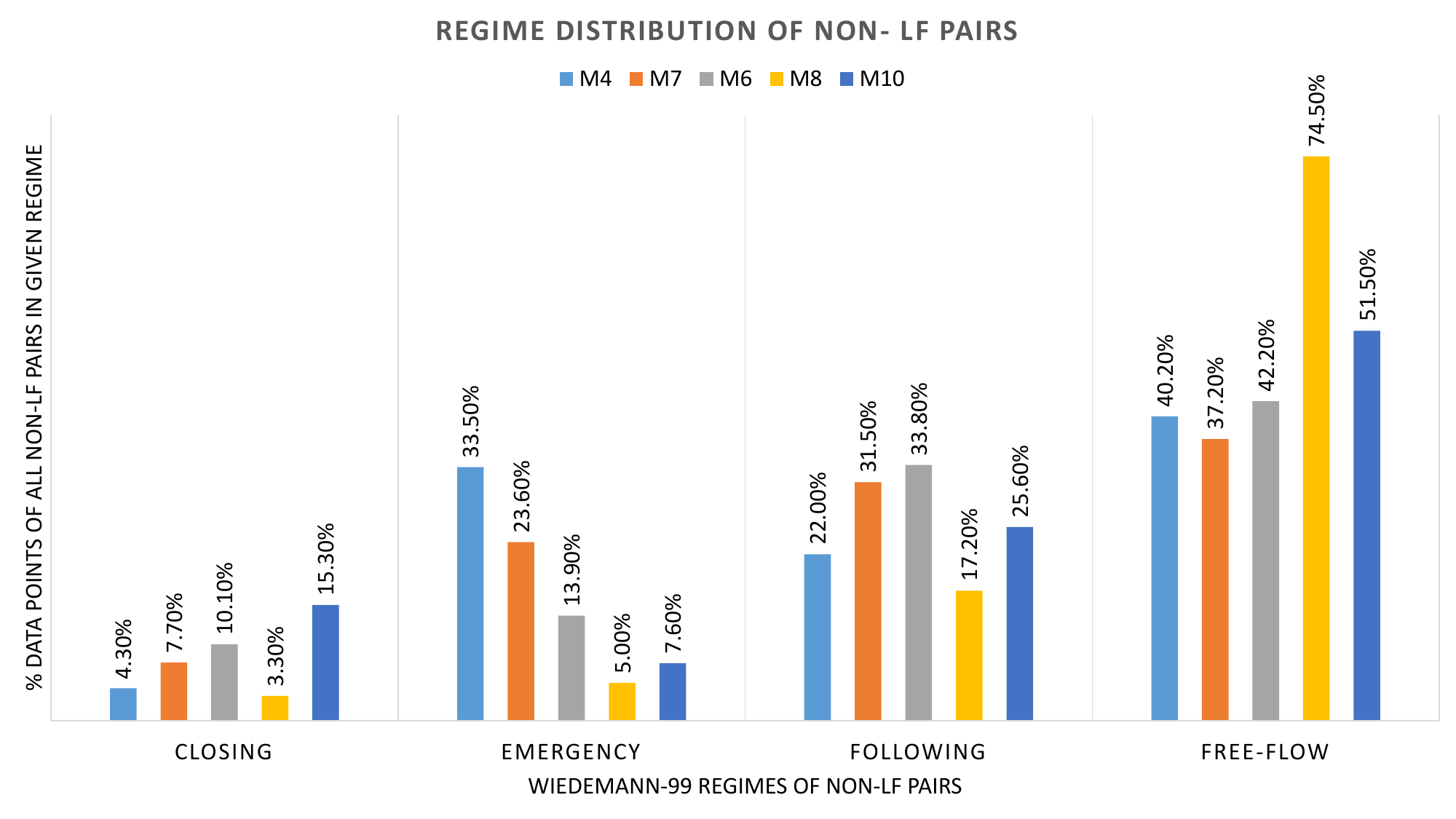} 
       \caption{}
    
      \end{subfigure}
  \caption{Wiedemann-99 regime distribution for LF pairs (a) and non-LF pairs (b) using different methods of LF identification}\label{Fig:regimedistLF}
\end{figure}

To demonstrate the advantage of the psychophysical concept in LF identification, we compared the regime distribution in LF and non-LF pairs for 5 methods as shown in Figure \ref{Fig:regimedistLF}. The fraction of points in the following regime of LF pairs (Figure \ref{Fig:regimedistLF} (a)) is the highest (67.20\%) for the proposed model (M8) which is an indicator of accurate LF identification. The second-best following regime fraction for LF pairs is for the headway model (M7), while other methods such as (M4) identified almost half of the points of LF pairs in the emergency braking regime, which leads to inaccurate prediction of trajectories as depicted with high GoF (Table \ref{tab:methods}). When regime distributions for non-LF pairs are compared (Figure \ref{Fig:regimedistLF} (b)), the proposed model (M8) has the highest fraction of free-flow regime (74.50\%), which ascertains non-following behavior in non-LF pairs. The free-flow regime fraction in non-LF pairs of the headway model (M7) is considerably less than that of the proposed model (M8). A relatively higher fraction following regime in non-LF pairs of the headway model (M7) and a lesser number of LF pairs (17.20\%) suggest that some of the pairs which are LF pairs, are identified as non-LF pairs by the headway model. Performance metrics and accurate LF identification results prove the superior performance of the proposed model (M8) compared to all the other methods.

\subsection{{Comparison of Methods Using Two-Dimensional Criteria}}
{While Methods M1-M7 only use longitudinal criteria, Methods M8-M13 make full use of the two-dimensionality of non-lane-based mixed traffic by defining lateral criteria and criteria for identifying intermediate vehicles.}

\begin{figure}[hbt!]
  \centering
  \includegraphics[width=1\textwidth]{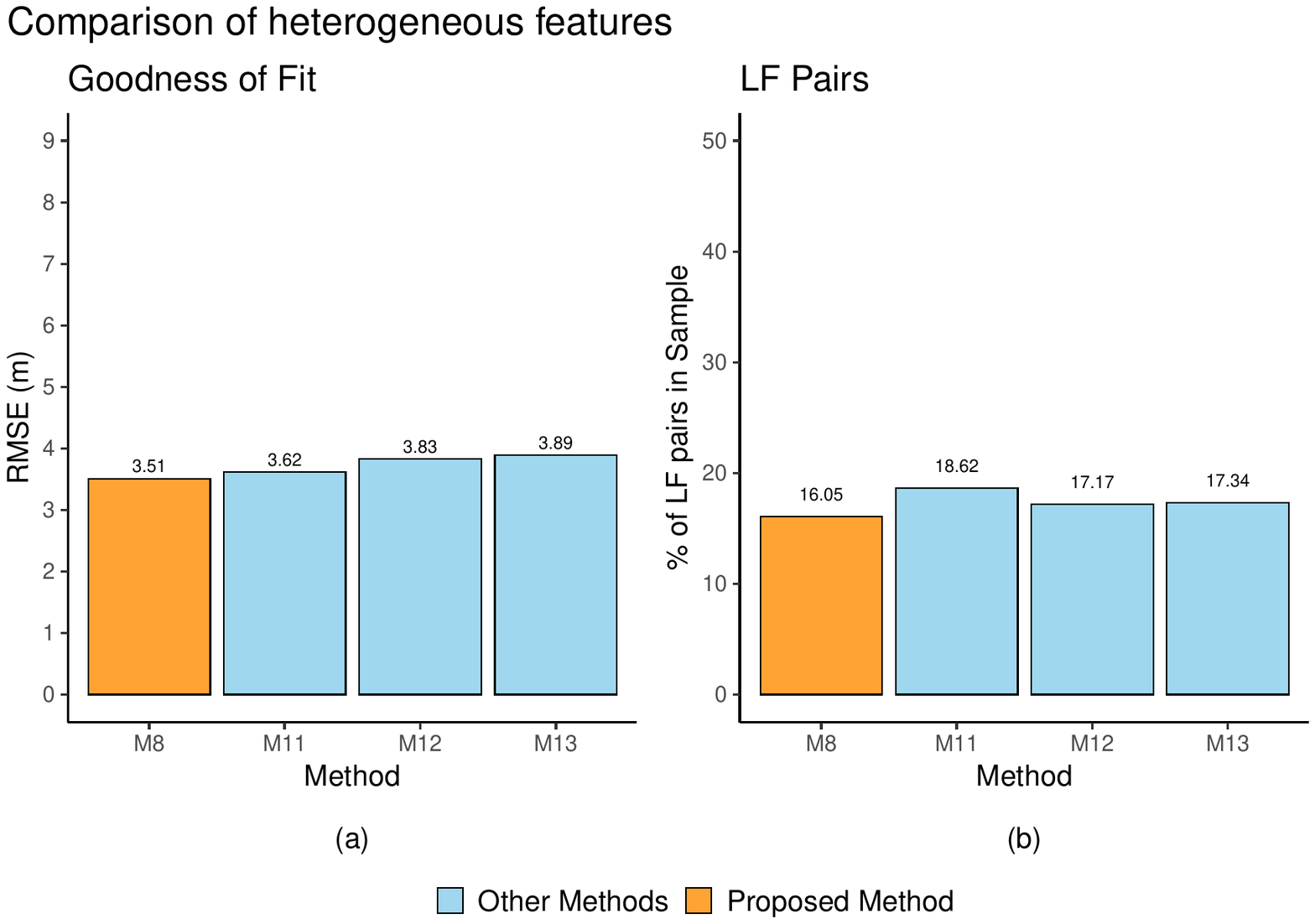}
  \caption{Comparison methods using two-dimensional criteria}\label{Fig:comparisonofhet}
\end{figure}

Figure \ref{Fig:comparisonofhet}(a) shows the comparison of {our method M8} with  Delaunay triangulation method M11 similar to \citep{nagahama2021detection} to identify an intermediate vehicle. If an edge of the triangulation appears between the assumed leader and follower, we say that the leader is influencing the follower at that instant. For {M11}, we consider the value of lateral clear gap threshold $c_0$ = 0.3 m as considered in \cite{nagahama2021detection}, {and with two variants M12 and M13 of our method using absolute and relative overlaps (negative gaps), respectively, instead of positive gaps.}  
As shown in Figure \ref{Fig:comparisonofhet}(b), a lesser variation is observed in the number of identified LF pairs. Consideration of lateral gap and intervening surrounding vehicles brings two important features of NLB traffic in LF identification and vehicle-following calibration. This result suggests that consideration of two-dimensional traffic features significantly improves model calibration. The proposed model, which does joint and consistent calibration, performs the best. Amongst the different lateral conditions, the superior performance of the lateral clear gap model suggests that when a car is a follower, positive overlap with the leader may not be required to have influence. Influence can be there even if there is a clear gap. The calibrated value of the threshold for the lateral clear gap is \unit[0.157]{m}. {Hence, by comparison between M8 and M12-M13 we can say that gaps are better suited to define thresholds than overlaps} 

\subsection{Statistical comparison of parameters using bootstrapping}

In this section, we conduct a comparative analysis of VF parameters obtained from bootstrapping using an unpaired t-test. As an example, we compare the parameters obtained from the proposed method (M8) with those from the headway method (M7). This procedure can be extended to all the methods. The aim of bootstrapping is to estimate the variance of each CC parameter {and to determine if the parameter values obtained from the two methods can be considered to be equal.} 

During the bootstrapping process, 100 pairs are selected with repetition from the LF pairs identified by the proposed method. Similarly, 44 pairs are selected from the headway method (M7). 
Notice that while the number of pairs in the sets is the same as the available number of LF pairs for the respective method, the sets are still different from the complete collection because, due to the selection with repetition, some pairs are drawn several times and some not at all. A hundred such sets are prepared for both methods. CC parameters are then calibrated for each set, and sample standard deviation \(s_1\) and  \(s_2\) are reported in Table 3. 
The pooled standard deviation \(s_p\) is calculated using (\ref{eq:sp}) and the standard error is calculated using (\ref{eq:SE}). The t-statistic for each parameter is calculated using (\ref{eq:T}). In this case, \(n_1\) = \(n_2\)= 100 since there are 100 sets for each method.

\begin{equation}\label{eq:sp}
    s_{\rm p} = \sqrt{\frac{(n_1-1){s_1}^2+(n_2-1){s_2}^2}{n_1+n_2+2}}
\end{equation}

\begin{equation}\label{eq:SE}
    SE(\bar{x_1}-\bar{x_2}) = s_p\sqrt{\frac{1}{n_1}+\frac{1}{n_2}}
\end{equation}

\begin{equation}\label{eq:T}
    T =\frac{\bar{x_1}-\bar{x_2}}{SE(\bar{x_1}-\bar{x_2})}
\end{equation}

 {The null hypothesis $H_0$ states that the investigated parameter has the same value for both methods.
 For CC1, CC2, CC3, CC4, CC5, and CC8, we can reject $H_0$ at error probabilities $p<\unit[5]{\%}$  while, for CC7 ($p=5.8\%$) there is not sufficient evidence for a rejection. 
 Notice that, during the bootstrapping investigation, we keep the LF parameters at the values of Table~\ref{tab:methods}.}

\begin{table}[!ht]
    \centering
    
    \caption{Statistical comparison of CC parameters using bootstrapping}
    \fontsize{10pt}{10pt}\selectfont
    \renewcommand{\arraystretch}{1.4}
    \resizebox{\textwidth}{!}
    {
    \begin{tabular}{lcccccccc}
    \hline
    \textbf{Method} & ~ & \textbf{CC1} & \textbf{CC2} & \textbf{CC3} & \textbf{CC4} & \textbf{CC5} & \textbf{CC7} & \textbf{CC8} \\
    \hline
    \multirow{2}{1.5em}{\textbf{Proposed (M8)}} & Sample mean (\(\bar{x}_1\)) & 0.303 & 6.799 & -3.707 & -2.439 & 2.408 & 0.665 & 3.062 \\
    ~ & Sample Std. dev. (\(s_1\)) & 0.074 & 1.818 & 2.385 & 0.467 & 0.449 & 0.176 & 0.400 \\ \hline
    
    \multirow{2}{5em}{\textbf{Headway (M7)}} & Sample mean (\(\bar{x}_2\)) &0.559 & 9.090 & -7.559 & -1.721 &1.916 &0.603 &2.843 \\
    & Sample Std. dev. (\(s_2\))  & 0.136 &1.929 &2.360 &0.459 &0.472 &0.213 &0.412 \\ \hline
    
    & Pooled Std. dev. (\(s_p\)) & 0.122 &2.273 &2.911 &0.569 &0.560 &0.232 &0.495 \\
    
    & Standard error (\(SE\)) & 0.017 &0.321 &0.412 &0.080 &0.079 &0.033 &0.070\\
    
    & t-statistic (\(T\)) & -14.907 &-7.126 &9.357 &-8.927 &6.223 &1.903 &3.134\\
    
    & Statistical  significance & Different & Different & Different & Different & Different & Same & Different \\ \hline
    
    \end{tabular}
    }
\end{table}

\subsection{Performance of validation set}
\begin{figure}[hbt!]
  \centering
  \includegraphics[width=1\textwidth]{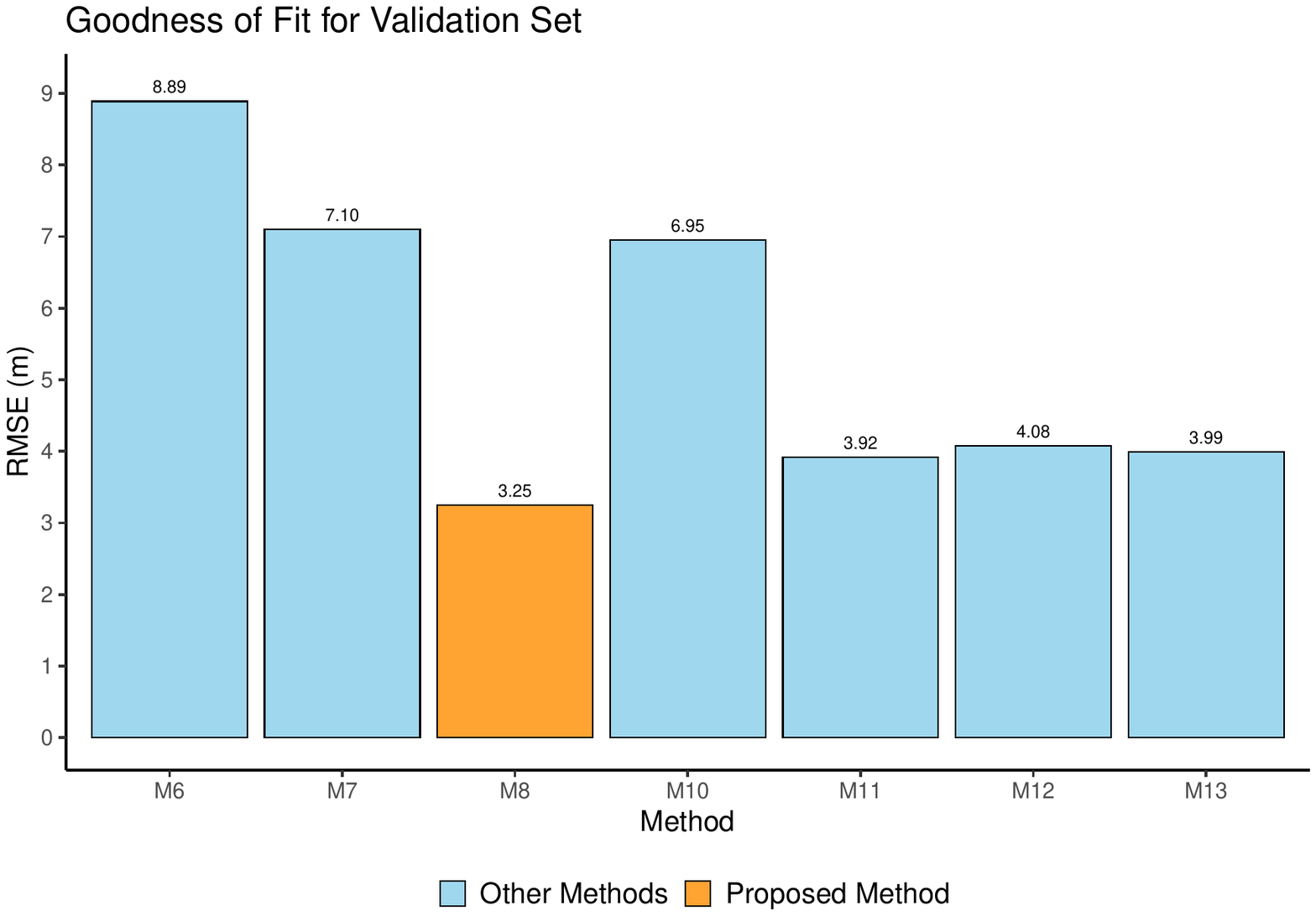}
  \caption{Performance of different methods on the Validation set}\label{Fig:validation}
\end{figure}

To validate the proposed methodology, it is necessary to assess its performance on unseen data. The available data consists of 623 potential LF pairs which are randomly divided into 70\% for estimation and the remaining 30\% for holdout validation. The estimation set contains 436 pairs, and the holdout set has 187 pairs. {We perform the validation on all joint and consistent methods} from Table \ref{tab:methods} (Figure \ref{Fig:validation}).
The proposed lateral clear gap (M8) method {clearly} shows the best results with the lowest RMSE position.

The results show that the proposed joint and consistent framework performs better for conventional heuristics as well as the proposed method. {While, on calibration, the methods M11-M13 nearly have the same fit quality as M8, there is a distinct difference in validation, showing the robustness of the proposed method.}

\section{Conclusion and Future Work}

In this paper, a novel method is proposed for jointly and consistently identifying LF pairs and calibrating vehicle-following model parameters in non-lane-based traffic flow. Existing LF identification methods are compared with the proposed method using empirical trajectory data from Chennai City, India. The proposed methodology leverages the concept of the regime in the psycho-physical Wiedemann-99 model to determine "influence zones". A logic to identify potential intervening vehicles between leader and follower was developed and used as one of the influencing conditions. Three LF identification parameters are introduced and calibrated, namely ‘minimum duration of the continuous following’ (\(t_{\rm cont}\)), ‘minimum fraction of influence points’ ($f_{\rm min}$) and ‘lateral gap threshold' $c_0$. Nested calibration was employed which iteratively calibrates CC parameters and LF identification parameters. The objective of the calibration was to minimize the mean of RMSE of position for LF pairs. 

The results{, particularly the validation,} demonstrate the superiority of the proposed joint and consistent calibration procedure compared to models that do not perform LF identification at all as well as those based on proximity or hysteresis-based heuristics. 

Overall, this study provides valuable insights into LF identification methods and their impact on vehicle-following behavior. The proposed methodology can be used to extract LF pairs from vehicle trajectory data under varying traffic conditions as well as using alternative vehicle-following models. In similar traffic conditions to our study area, the calibrated Wiedemann-99 car-following parameters and LF identification parameters like lateral clear gap, can be directly adopted to obtain LF pairs. Lateral thresholds and continuous following duration thresholds could be used to further investigate lateral (lane-changing) behavior of heterogeneous and non-lane-based traffic which is an insufficiently explored research area. 

This research has implications for traffic engineering, as the proposed approach can enable robust and more realistic LF identification and VF parameter calibration. The incorporation of these techniques within a microsimulation framework can yield more accurate traffic fundamental diagrams (FDs) for non-lane-disciplined traffic, which has a wide range of applications including LOS, capacity, and travel time analysis. With a better understanding of vehicle-following behavior, highway capacities can be estimated accurately and FDs could be used to determine reliable travel time estimates. In our next research thread, we aim to develop FDs using the traffic simulation packages.

There are several directions in which this work could be extended. Currently, we only analyzed the behavior of the cars as a follower, but two-wheelers are also significant contributors to the traffic volume in the study area. Investigating the influence conditions for two-wheelers would improve traffic simulation. Moreover, our proposed model focuses on vehicle-following behavior irrespective of the type of leader, but, the following behavior could be impacted by the vehicle type of the leader. The proposed methodology could be utilized to understand the following behavior of various LF types. The evaluation of alternative vehicle-following models within this framework is yet another promising avenue of research. Also, we have only considered the longitudinal aspects (VF), and future work includes developing a trajectory-based calibration methodology for the fully 2D behavior, e.g., for the Intelligent-Agent Model \citep{treiber2023intelligent}. 

\section*{Declaration of competing interest}
The authors declare that they have no known competing financial interests or personal relationships that could have appeared to influence the work reported in this paper.

\section*{Acknowledgments}
The work is funded by a scholarship from the Ministry of Education, Government of India. DAAD (The German Academic Exchange Service), KOSPIE fellowship. And DFG (Deutsche Forschungsgemeinschaft) under project grant no. 456691906, "Enhancing Traffic Flow Understanding by Two-Dimensional Microscopic Models - ETF2D" at TU Dresden. The authors also acknowledge \cite{kanagaraj2015trajectory} for sharing the trajectory dataset.

\section*{CRediT authorship contributions statement}
Mihir Mandar Kulkarni: Conceptualization, Methodology, Investigation, Writing – original draft, Writing – review \& editing. Ankit Anil Chaudhari: Conceptualization, Methodology, Investigation, Writing – original draft, Writing – review \& editing. Karthik K Srinivasan: Conceptualization, Methodology, Writing – review \& editing. Bhargava Rama Chilukuri: Conceptualization, Methodology, Writing – review \& editing. Martin Treiber: Conceptualization, Methodology{, Writing review \& editing.} Ostap Okhrin: Conceptualization, Methodology, Writing – review \& editing.  All authors reviewed the results and approved the final version of the manuscript.

\section*{Declaration of generative AI and AI-assisted technologies in the writing process}

During the preparation of this work, the author(s) used ChatGPT in order to do grammar checks and language corrections. After using this tool/service, the author(s) reviewed and edited the content as needed and take(s) full responsibility for the content of the publication.

\appendix

\section{Wiedemann-99 Model Regime Classification }
The equations for calculating Wiedemann-99 regimes based on vehicle-following parameters are as follows, \citep{anil2022calibrating}:

\begin{align*}
    AX &= CC0, 
    &ABX &= CC0 + CC1 \cdot V_{\text{slow}}, \\
    SDX &= ABX + CC2,
    &CLDV &= CC5 + \frac{CC6}{17000} \cdot DX^2, \\
    OPDV &= CC4 - \frac{CC6}{17000} \cdot DX^2,
    &SDV &= CC5 + \frac{DX-SDX}{CC3},
\end{align*}
where
\begin{itemize}
    \item AX: the desired distance between two vehicles in a stopped condition;
    \item ABX: the desired minimum safe following distance in moving state, as a lower limit of the following regime;
    \item SDX: the maximum following distance as the upper limit of the following regime;
    \item CLDV: the points at short distances (less than SDX) where drivers perceive that their speeds are higher than their lead vehicle speeds;
    \item OPDV: the points at short distances (less than SDX) where drivers perceive that they are traveling slower than their leader;
    \item SDV: the points at long distances (more than SDX) where drivers perceive that they are approaching slower vehicles.
\end{itemize}
Wiedemann-99 car following parameters are explained as follows:
\begin{itemize}
    \item CC\(0\) [\unit[0.65]{m}]: The desired gap between two vehicles in a stopped condition;
\item CC\(1\) [\unit[0.9]{s}]: Time gap the following driver keeps for a safety-in-moving state;
\item CC\(2\) [\unit[4]{m}]: Range of the gap between vehicles in the following regime;
\item CC\(3\) [\unit[-8]{s}]: The time between the beginning of deceleration after perceiving a slow-moving leader and starting the unconscious following behavior;
\item CC\(4\) [\unit[-0.35]{m/s}]: Speed difference during the following process— CC\(4\) controls speed differences during the opening process (negative relative speed);
\item CC\(5\) [\unit[0.35]{m/s}]: Speed difference during the following process— CC\(5\) controls speed differences in the closing process (positive relative speed);
\item CC\(6\) [\unit[11.44]{1/ms}]: Influence of distance on speed oscillation during the following condition;
\item CC\(7\) [\unit[0.25]{m/$s^2$}]: Actual acceleration during oscillation in the unconscious following regime;
\item CC\(8\) [\unit[3.5]{m/$s^2$}]: Desired acceleration when the vehicle starts from the standing condition.
\end{itemize}

 \bibliographystyle{elsarticle-harv} 
 \bibliography{ref}

\end{document}